\documentclass{caps}
\usepackage{epsf}

\begin{document}

\setcounter{chapter}{14}

\pagenumbering{roman}
\tableofcontents
\cleardoublepage

\pagenumbering{arabic}

\author[K. Hurley, R. Sari, and S. G. Djorgovski]{K. HURLEY\\University of California\\Space Sciences Laboratory\\Berkeley, CA
\and R. SARI\\California Institute of Technology\\Theoretical Astrophysics\\Pasadena, CA
\and S. G. DJORGOVSKI\\California Institute of Technology\\Palomar Observatory\\Pasadena, CA}

\chapter[Gamma-ray bursts]{Cosmic Gamma-Ray Bursts, Their Afterglows, and Their Host Galaxies}

\vspace{4cm}

\section{Introduction}

Regarded as an astrophysical mystery and a curiosity for decades, cosmic gamma-ray
bursts are finally entering the mainstream of astronomy and astrophysics.  In the past five
years, we have learned that they lie at cosmological distances, and are probably caused,
possibly among other things, by the collapses and subsequent explosions of massive stars.
Energetically they are roughly analogous to supernovae, to which they may indeed be
related in some cases; no new physics needs to be invented to explain their prodigious
luminosities.  Unlike supernovae, however, they are relatively rare, and their energy output is
distributed quite differently over wavelength and time.  They can probably be observed
out to distances comparable to, or even farther than, those of the most distant quasars,
which makes them useful to cosmologists as lighthouses to the early universe.  Finally, too,
they hold the promise of revealing properties of early galaxies such as star formation
rates and metallicities in ways that are unique.  For all of these reasons, in addition to the
facts that they signal the formation of black holes and drive ultra-relativistic winds, they
have begun to attract the attention of people working in very diverse disciplines.  The words
``gamma-ray burst'' have even begun to enter the vocabulary of the general public, which
regards them with a certain morbid fascination.

It was not at all clear a decade ago that the study of gamma-ray bursts (GRBs) had
such a promising future.  If, as many people then suspected, they were generated by
some sort of activity on galactic neutron stars, they would probably not have been
observable for more than the 10's of seconds of the bursts themselves, and they
might well have remained a curiosity.  Two popular accounts of how our understanding of GRBs evolved have
now appeared  (Katz 2002; Schilling 2002), so in this chapter we will forego the intriguing history
of the subject, and begin by describing the phenomenology of bursts.  The brief gamma-ray
emitting phase is followed by a longer duration, long-wavelength ``afterglow''; the characteristics
and the theory of afterglows are described next.  Finally, the observations of afterglows often
lead to the identification of the host galaxies of bursts, which are treated in the last section
of this chapter.   

In this rapidly evolving field, some of the most up-to-date information is found
on websites.  A non-exhaustive list of them is:\\

\noindent The Interplanetary Network: ssl.berkeley.edu/ipn3/index.html \\
HETE-II: space.mit.edu/HETE \\
BATSE:  www.batse.msfc.nasa.gov/batse/ \\
BeppoSAX: www.asdc.asi.it/bepposax/ \\
Swift: swift.gsfc.nasa.gov \\
The gamma-ray burst coordinates network: gcn.gsfc.nasa.gov/gcn/ \\
Jochen Greiner's afterglow website: www.mpe.mpg.de/$\sim$jcg/grb.html \\
A radio catalog of gamma-ray burst afterglows: www.aoc.nrao.edu/$\sim$frail/grb\_public.html

The most recent conference proceedings is ``Gamma-Ray Bursts in the Afterglow Era'' , edited
by Costa, Frontera, \& Hjorth (2001).

\section{The big picture}

A typical GRB occurs in a star-forming region of a galaxy at a redshift z $\approx$ 1.  In currently
popular models, it is caused by the collapse of a massive star ($\approx$ 30 solar masses) which
has exhausted its nuclear fuel supply.  The star collapses to a black hole threaded by a strong
magnetic field, and possibly fed by an accretion
torus.  In this configuration, energy can be extracted through the Blandford-Znajek (1977) mechanism.
This energy goes into accelerating shells of matter, once part of the massive star, to ultra-relativistic
velocities (Lorentz factors of several hundred).  These shells collide with one another as they
move outward, producing ``internal'' shocks in a solar-system sized volume.  The shocks accelerate
electrons, and the electrons emit synchrotron radiation.  In the observer's frame, the radiation appears
in gamma-rays, and produces a burst with $\approx$ 20 second duration.  If the gamma-rays were
emitted isotropically, they would account for well over 10$^{53}$ erg of energy in many cases.
However, there is evidence that this gamma-radiation is
strongly beamed, within a cone whose opening angle is only several degrees (Frail et al. 2001) and thus
that the total energy emitted in this stage is some two orders of magnitude smaller.  As the shells continue
to move outward, they eventually reach a region of enhanced density.  This could be either the interstellar
medium, or a region which was populated with matter by the massive star in its final stages of evolution.
As the shells impinge on this region, they produce ``external'' shocks, which give rise to a long-lived radio,
optical, and X-ray afterglow which may be detectable for years in the radio, weeks to months in the optical,
and weeks in X-rays.  There is about an order of magnitude less energy in the afterglow than
in the burst itself.  Initially, this afterglow radiation is beamed, but as the shells decelerate, they spread
laterally and the radiation tends towards isotropy. The afterglow tends to fade as a power law with time.  
However, in many cases, the decline is not
completely monotonic; ``bumps'' can appear in the optical lightcurve, and they have been interpreted either as
a supernova-like component or as the result of microlensing. 

The model described above is known as the ``standard fireball model''.  Such models had been discussed
extensively long before the GRB distance scale was known, but the establishment of a cosmological
distance scale for bursts brought them into sharp focus (Wijers, Rees, \& Meszaros 1997).  To be sure, there are competing models,
as well as variations on this theme, and they cannot be ruled out.  Afterglows are only detected for about
one-half the bursts.  In those cases where they are not detected, the host galaxies cannot be identified,
and it is almost impossible to demonstrate that the GRB is due to the collapse of a massive star, as
opposed to the merger of two neutron stars, for example.

Because the gamma-rays are beamed, we detect only a small fraction of them.  The most sensitive
GRB detector flown (BATSE aboard the Compton Gamma-Ray Observatory) detected roughly one
burst per day down to its threshold, and missed about one per day due to well understood effects such as Earth-blocking.
Using current estimates of beaming, this implies that the Universe-wide GRB rate is at least 1000/day,
and possibly more if there are many weaker bursts that were not detected by BATSE. 

\section{Some technical details}

Before the radio, optical, or X-ray afterglow can be identified, the burst must be localized rapidly (i.e. within a
day or so) to reasonable
accuracy (several 10's of arcminutes) during the bursting phase, which typically lasts only several 10's of seconds.
The two main ways to do this are first, using a coded mask detection system (e.g. BeppoSAX, Costa et al. 1997) and second, by timing
the arrival of the burst at spacecraft separated by interplanetary distances (the Interplanetary Network, Hurley et al. 2000).  Burst detection rates
using these techniques can reach $\approx$ 0.5 to one per day.

Next, the position of the burst must be communicated rapidly to observers.  This is now done almost exclusively
through the Gamma-Ray Burst Coordinates Network (GCN, Barthelmy, Cline, \& Butterworth 2001), which reaches almost 600
recipients.  In the early phases, the afterglow may be bright enough to be detected by amateur astronomers with
telescopes of modest size (m $\approx$ 16 in the day or so following the burst; in many cases, the burst
outshines its host galaxy).  
This phase is crucial, because the position of an optical afterglow must be determined
to arcsecond accuracy or better to allow optical spectroscopy to take place with telescopes of much larger size, as
well as to permit deep observations which may reveal the presence of a host galaxy.  

In many cases, the X-ray afterglow has been observed within hours of the burst by the same spacecraft which localized the burst,
after slewing to the position (BeppoSAX, Costa 2000).  In other cases, target-of-opportunity observations have been
carried out with a different spacecraft from the one which detected the burst, but with much longer delays.

Radio measurements can be carried out at a more leisurely pace, since the radio emission tends to peak
days after the event (Frail, Waxman, \& Kulkarni 2000).  But only the largest radio telescopes need apply: typical peak fluxes
are at the milliJansky level.    

Over the years, astronomers have become extremely adept at identifying GRB counterparts, to the point
where redshifts have been measured in less than 8 hours from the time of the burst.

\section{The bursting phase}

\subsection{Gamma-ray burst lightcurves}

\begin{figure}
\centering
\epsfxsize30pc   
\epsfbox{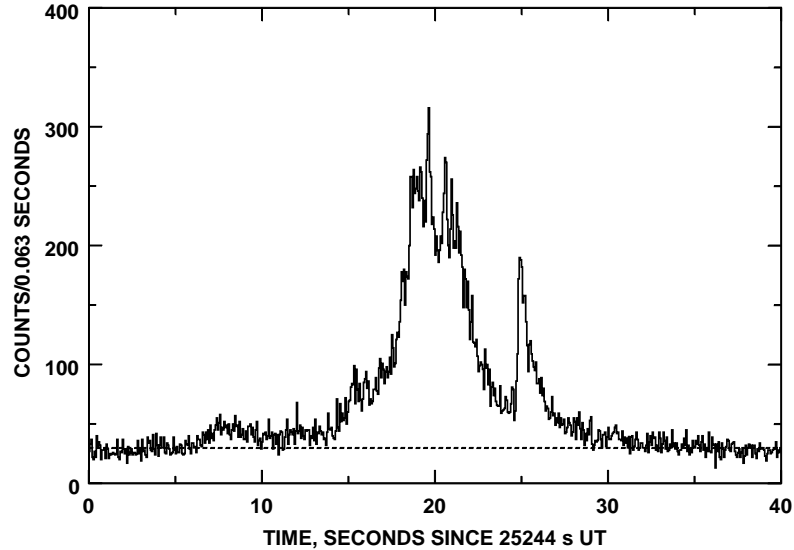}
\caption{A bright burst observed by the Ulysses GRB experiment on October
8 2002.  The energy range is 25-150 keV.  The dashed line indicates the background level.  The 25-100 keV fluence of
this event was $\rm \approx 8 \times 10^{-5} erg/cm^2$.}
\label{figure1}
\end{figure}

Gamma-ray bursts are, for a few seconds, the brightest objects in the gamma-ray sky.
Figure ~\ref{figure1} shows an example.
Indeed, bursts are so bright that an uncollimated, unshielded detector with a surface area of only 20 cm$^{2}$ can
detect a burst out to a redshift of z=4.5 (Andersen et al. 2000).

Burst durations span about 5 orders of magnitude, from 0.01 to 1000 s.  The duration distribution displays
a clear bimodality, with short bursts (durations $\approx$ 0.2 s) comprising around 25\% of the total,
and long bursts (durations $\approx$ 20 s) comprising the remainder (Mazets et al. 1981b; Dezalay et al. 1996;  Norris et al. 1984;  Hurley 1992;
Kouveliotou et al. 1993).  The distribution may be described by a lognormal function (Mcbreen et al. 1994).  Apart from this, gamma-ray burst lightcurves 
are generally very different from one burst to another, although certain
morphological types have been noted.  For example, about 7\% of all bursts display
a fast-rise, exponential decay morphology (Bhat et al. 1994).  What determines the shape of the lightcurve in this and
other morphologies is unknown.

In the lightcuves of very intense bursts, it is sometimes possible to detect a long, faint tail after the intense
emission has ceased (Burenin et al. 1999).  The count rate in the tail falls as a power law with time.  This is a relatively
short-lived ($\approx$ 1000 s) gamma-ray afterglow.  There is also evidence that the X-ray afterglow starts
during the gamma-ray burst in many cases (Frontera et al. 2000).

\subsection{Energy spectra}

GRB energy spectra have been measured from $\approx$ 2 keV (Frontera et al. 2000) to 18 GeV
(Hurley et al. 1994).  Even at the highest energies accessible to spark chamber detectors, there is
little or no evidence for spectral breaks.  Indeed, there is even tantalizing evidence for TeV
emission from one burst (Atkins et al. 2000).  The spectra may be fit over a wide range with various
models.  One is the so-called ``Band model'' (Band et al. 1993) which has no particular physical
derivation. Another is a synchrotron spectrum (Bromm \& Schaefer 1999).  An example of the latter is
shown in figure 2.  The spectrum in this figure is plotted in $\rm \nu F_{\nu}$ units, which
make it clear that the peak of the energy output  during the burst, $\rm E_{peak}$, is indeed at gamma-ray energies.
The distribution of $\rm E_{peak}$ derived from BATSE data is quite narrow (Mallozzi et al. 1995), which is surprising considering
the great diversity exhibited by most other GRB characteristics.  The extent to which the $\rm E_{peak}$
distribution could be biased due to detector characteristics has been considered.  There is presently
no compelling evidence that a population of very high $\rm E_{peak}$ GRBs exists (Harris and Share 1998),
although not all the phase space has been searched for such events.  On the other hand, however,
there \it is \rm evidence for one or two classes of soft-spectrum bursts, called ``x-ray rich GRBs'' and
``x-ray flashes'' which appear to have all the characteristics of gamma-ray bursts, except for the gamma-
rays above 20 keV or so (Heise et al. 2001). These may account for up to 30\% of the total bursts.  Some
of these were in fact detected by BATSE (Kippen et al. 2001), but it seems plausible that many were not,
leading to a possible bias against these events.

\begin{figure}
\centering
\epsfxsize20pc   
\epsfbox{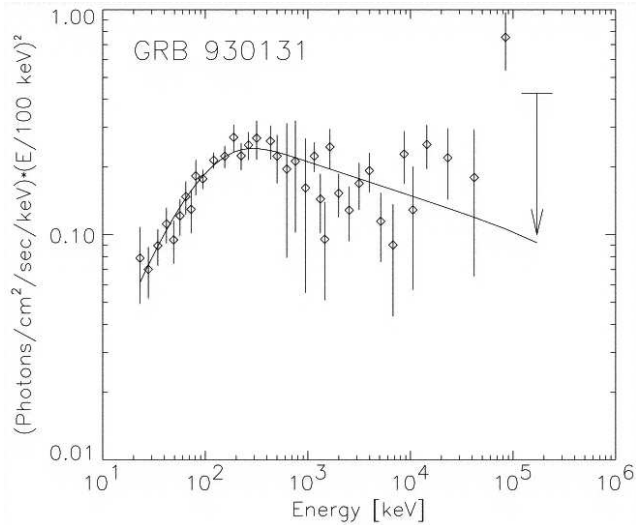}
\caption{The spectrum of GRB 930131 over four
decades, and a synchrotron model fit (Bromm \& Schaefer 1999).  The spectrum is plotted in $\rm \nu F_{\nu}$ units,
which show the amount of energy per decade.  The peak is at an energy $\rm E_{peak}\approx 200 keV $,
and the energy output declines only slightly at higher energies.}
\label{figure2}
\end{figure}

An interesting correlation exists between the time histories and the energy spectra of bursts:
the short bursts have harder energy spectra than the long bursts (Dezalay et al. 1996; Kouveliotou et al. 1993).
It is presently not understood why this is the case.

\subsection{GRB statistics}

It has been known for a long time that the spatial distribution of bursts is isotropic, to varying
degrees of statistical uncertainty e.g. (Mazets et al. 1981a).  It has similarly been known that the GRB number-
intensity relation, or log N-log S curve, displayed a turnover at low intensities, that is, a paucity of
weak events with respect to the -3/2 power law expected for a homogeneous distribution in
Euclidean space (Mazets et al. 1981b).  However, it was not until BATSE results became available that
the situation was put into perspective and clarified (Pendleton et al. 1996; Paciesas et al. 1999).  With 
unprecedented statistics and careful attention to instrumental details, the BATSE results confirmed
the isotropy of bursts.  They also confirmed the turnover in the log N-log S curve, albeit at much
smaller intensities than before, indicating
that the earlier turnover was due to instrumental effects (mainly the loss of sensitivity to weak
events).  The present results are consistent with a cosmological population of bursts (e.g. Stern, Atteia, \& Hurley 2002).
By some estimates, GRBs may occur out to redshifts of 10 or more, and current missions such as
HETE are capable of detecting them to z=8 (Lamb and Reichart 2000).

It is obviously interesting to sample the GRB population below the BATSE trigger threshold.  
For example, if there were an epoch of early star formation in the Universe which
gave rise to GRB-producing stellar deaths, this might be manifested by an increase in the
log N-log S curve at low intensities.   
Two recent studies have succeeded in exploring the low intensity population (Kommers et al. 2000; Stern et al. 2001).  By identifying bursts that
were too weak to trigger the BATSE detector, they have effectively reduced the threshold by a factor
of about 2.  However, the conclusions of the studies differ; in one case, there appears to be evidence that
the curve continues to rise at low intensities, suggesting that bursts continue to originate from earlier and
earlier parts of the Universe, while in the other case, there is evidence for a flattening of the curve.
It will require the next generation of GRB detectors to sort this out.

\subsection{Burst types and classes}

There are many \it types \rm of gamma-ray bursts.  They are often referred to as \it classes \rm, but it is
not known whether they actually originate from different kinds of explosions, or the explosions of different
kinds of stars, as opposed to say, originating from different viewing angles or other observing conditions.
A brief, non-exhaustive summary of burst types follows.

  \begin{enumerate}\listsize
 \renewcommand{\theenumi}{(\alph{enumi})}

\item Long and short bursts.  The duration distribution is bimodal, and the energy spectra of the short
bursts are harder than those of the long bursts.  No radio, optical, or X-ray counterpart has been found
for any short event (Hurley et al. 2002).  It has been speculated that the short bursts might arise from neutron star-
neutron star mergers (e.g. Macfadyen and Woosley 1999a), which could take place far from a host galaxy, and lack an
ISM on which to produce a long-lived afterglow.

\item Dark bursts.  While virtually all long bursts display X-ray afterglows, only about one-half of them
have detectable radio or optical afterglows.  There are various ways to hide intrinsically bright afterglows
and make them undetectable, such as beaming them away from the observer, absorbing the
light in the host galaxy, and placing the burst at high redshift.  Yet another is to invoke a flat spectral
shape (e.g. Hjorth et al. 2002).  It is possible that more than one explanation is required.

\item Bursts possibly associated with supernovae.  The first such event was GRB980425 (Galama et al. 1998c),
a burst whose position and time of occurrence were both consistent with those of an optical supernova, 1998bw.  In
other cases, supernova-like bumps in the afterglow light curves have been identified and attributed to underlying
supernovae (Bloom et al. 1999).  In still other cases, there is no evidence for such components.

\item X-ray flashes (XRF's).  These are bursts which resemble GRBs in almost every respect: durations, spatial
distributions, etc.  However, they display little or no emission above $\approx$ 25 keV (Heise et al. 2001).  Possibly
related to them are the X-ray rich GRBs, which display some gamma-ray emission (there is no widely accepted
definition yet for just what ratio of X-ray flux to gamma-ray flux constitutes an XRF).  One way to eliminate gamma-
radiation is to redshift the burst.  However, one X-ray rich GRB, 021004, has a redshift of only 1.6 (Fox et al. 2002), so
this cannot be the only explanation.

 \end{enumerate}

\section{GRB theory - the generic picture \label{sec:grbtheo}}

The mechanism leading to the phenomenon of GRBs is yet a matter of
debate. Nevertheless, some basic characteristics are well understood.
Below, we show how the observed spectra, energies and timescales of
GRBs have led to a generic model, the so-called fireball shock model
that is almost independent of knowledge about the unknown `inner
engine'.

The extreme characteristics of GRBs, i.e. the observed large energies
and short timescales, lead to a paradox, the `compactness
problem'. An energy of $10^{52}$ erg is released within a variability
time $\delta T \sim 0.1$s in the form of $\approx$ 1 MeV photons.
This translates into a huge number of photons,$N=10^{56}$. If we
now assume that the energy is released in a small volume of linear
dimension $R\le c\delta T \sim 3\times 10^{9}$cm (which is naively
required by the variability timescale), then the optical depth to pair
creation would be the number of photons per unit area, multiplied by
the Thomson cross section $\sigma_T$ or
$$
\tau \sim \sigma_T {N \over 4 \pi R^2} \sim 3 \times 10^{11} \gg 1.
$$ 
But, if that were true, it would imply that all the
photons will have created pairs and thermalized. However,
the observed spectrum of GRBs, as shown in the previous section is
highly non-thermal!

\begin{figure}
\vspace{-50pt}
\epsfxsize40pc 
\epsfbox{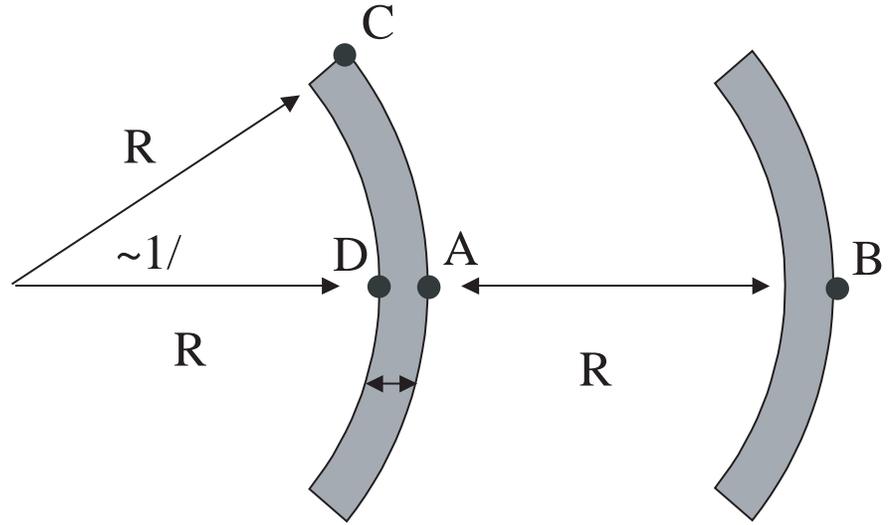}
\vspace{-250pt}
\caption{ Timescales from an expanding relativistic fireball. The
gray area represents the observed section of the fireball that can be
seen by an observer located far to the right. The angular opening of
that section is $1/\gamma$ due to relativistic beaming. Consider the 4
photons emitted at points A, B, C, and D. Photons A, C and D where
emitted simultaneously, but photon A will arrive at the observer
first, since it is closer to the observer.  The arrival-time delay of
photons C and D with respect to photon A is simply given by the extra
distance they have to travel. Therefore $\delta T_{C-A}=R(1-\cos
\theta)/c = R/2\gamma^2c$, and $\delta T_{D-A}=\Delta/c \sim
R/\gamma^2 c$, where we have used the fact that relativistic dynamics
of fireballs imply $\Delta \sim R/\gamma^2$. Finally, photon B was
emitted long after photon A (about a time $R/c$ later than photon A);
however, it is much closer to the observer, resulting in $\delta
T_{B-A} = R/2 \gamma^2 c$. All three timescales lead to the expression
$R/\gamma^2 c$. A short observed variability time scale can therefore
be obtained even for large radius, if the Lorentz factor is
sufficiently high. The naive estimate of $R \le c \delta T$ is,
therefore, to be replaced by $R \le \gamma^2 c \delta T$.
\label{fig:timescales}
}
\end{figure}

The only known solution to the `compactness problem' is relativistic
motion (Paczy\'nski 1986; Goodman, 1986). These effects have been considered in detail
(Krolik \& Pier 1991; Fenimore, Epstein, \& Ho 1993; Baring \& Harding 1997). A critical review of these as well as some new
limits are given by Lithwick and Sari (2001). If the emission
site is moving relativistically toward the observer with a Lorentz factor $\gamma$,
then the optical depth is reduced compared to
the stationary estimate, due to two effects. First, the size of the
source can be larger by a factor of $\gamma^{2}$. This will still
produce variability over a short time scale given by $\delta
T=R/\gamma^{2}c$ since not all of the source is seen because the radiation
for a relativistically moving object is beamed (see figure
\ref{fig:timescales}). Second, the photons in the local frame are
softer by a factor of $\gamma$, and therefore only a small fraction of
them, the ones at the high-energy tail of the GRB spectrum, have
enough energy to create pairs. The combination of these two effects
reduces the optical depth by a factor of $\sim \gamma ^{6.5}$, where
the exact power depends on the GRB spectrum (see
Lithwick \& Sari 2001). Therefore, the optical depth is reduced below unity, and
the `compactness problem' is solved, if the Lorentz factor is larger
than about one hundred.

This solution to the compactness problem led to a three stage generic
scenario for GRBs. First, a compact source releases about $10^{52}$
erg, in a small volume of space and on a short time scale. This large
concentration of energy expands due to its own pressure (M\'esz\'aros \& Rees 1993;
Piran \& Shemi 1993; Piran, Shemi, \& Narayan 1993). If the rest
mass that contaminates the site is not too large, $\le 10^{-5}M_{\odot
}$ (the requirement of a small baryonic load), this will result in
relativistic expansion with $\gamma>100$. Finally, at a large enough
radius, the kinetic energy (bulk motion) of the expanding material is
converted to internal energy and radiated, mainly in $\gamma$-rays. At
this stage the system is optically thin and high energy photons can
escape. We now discuss this third stage in some detail.

\subsection{Internal vs. external shocks \label{sec:intext}}

Assume a flow carrying $10^{52}$ erg as kinetic energy. In order for
this to produce photons, the kinetic energy must be converted back
into internal energy and radiated away. The flow must therefore, at
least partially, slow down. Two scenarios were proposed for this
deceleration: external shocks (M\'esz\'aros \& Rees 1993) and internal shocks
(Narayan, Paczy\'nski, \& Piran 1992; Ress \& M\'esz\'aros 1994). In the external shocks scenario, the relativistic
material is running into some (external) ambient medium, possibly the
interstellar medium (ISM) or a stellar wind that was emitted earlier by the
progenitor. In the internal-shocks scenario the inner engine is
assumed to emit an irregular flow consisting of many shells that
travel with a variety of Lorentz factors and therefore collide with
one another and thermalize part of their kinetic energy.

The property that proved to be very useful in constraining these two
possibilities is the variability observed in many of the bursts. In
the external shocks scenario, this variability is attributed to
irregularities in the surrounding medium, e.g., clouds. Each time the
ejecta run into a higher density environment, they produce a peak in
the emission. In the internal shocks scenario, the source has to emit
many shells, and when two of them collide a peak in the emission is
produced. External shocks thus require a complicated surrounding with
a relatively simple source that explodes once, while internal shocks
require a more complicated source that will explode many times to
produce several shells. Due to these very different requirements on
the source, the question of internal or external shocks is of
fundamental importance in understanding the nature of the
phenomenon.

The size of the clouds that the ejecta run into, in the
external-shocks scenario, has to be very small in order to produce
peaks that are narrower than the duration of the burst
(Fenimore, Madras, \& Nayakchin 1996). Sari \& Piran (1997a) gave the following argument.
The size of the clouds has to be smaller than $R/N\gamma $ to produce
peaks that are narrower by a factor of $N$ than the duration of the
burst. The number of clouds should be smaller than $N$ otherwise
pulses arriving from different clouds will overlap and the amplitude
of the variability will be reduced. Finally,due to relativistic beaming, the observable area of the
ejecta  is $(R/\gamma)^{2}$. The maximal
efficiency of the external shocks scenario is therefore given by
\begin{equation}
\frac{\mathrm{cloud\ area \times number\ of\
clouds}}{\mathrm{observed\ shell\ area}}\le \frac{1}{N}\sim1\%.
\end{equation}
Since in many bursts $N>100$, external shocks have a severe efficiency
problem in producing highly variable bursts.  The problem is even more
dramatic if long quiescent periods, which are observed in many bursts
(Nakar \& Piran 2002), are taken into account.  Also, other
predictions of external shocks are inconsistent with the observed
temporal profile (Ramirez-Ruiz \& Fenimore 1999). Moreover, the density ratio between the
clouds and their surroundings has to be huge, of the order of $\gamma
N^{2}\sim 10^{6}$, in order for the ejecta to be slowed down mainly
by the dense clouds rather than by the low density medium that they
are embedded in.  Finally, we mention that despite the above arguments, some still
favor other scenarios (Dermer \& Mitman 1999; Dar \& DeRujula 2000).

\begin{figure}

\centerline{\epsfxsize15pc\epsfbox{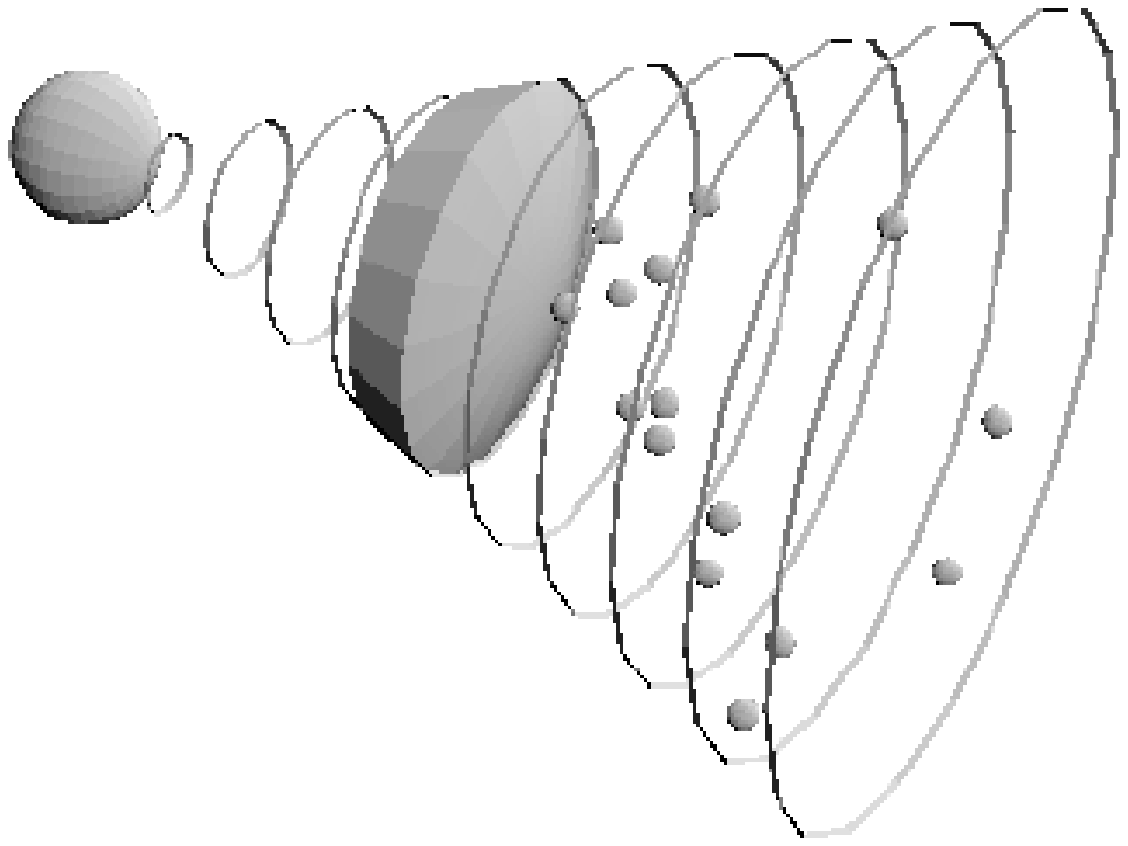}\ \ \epsfxsize15pc\epsfbox{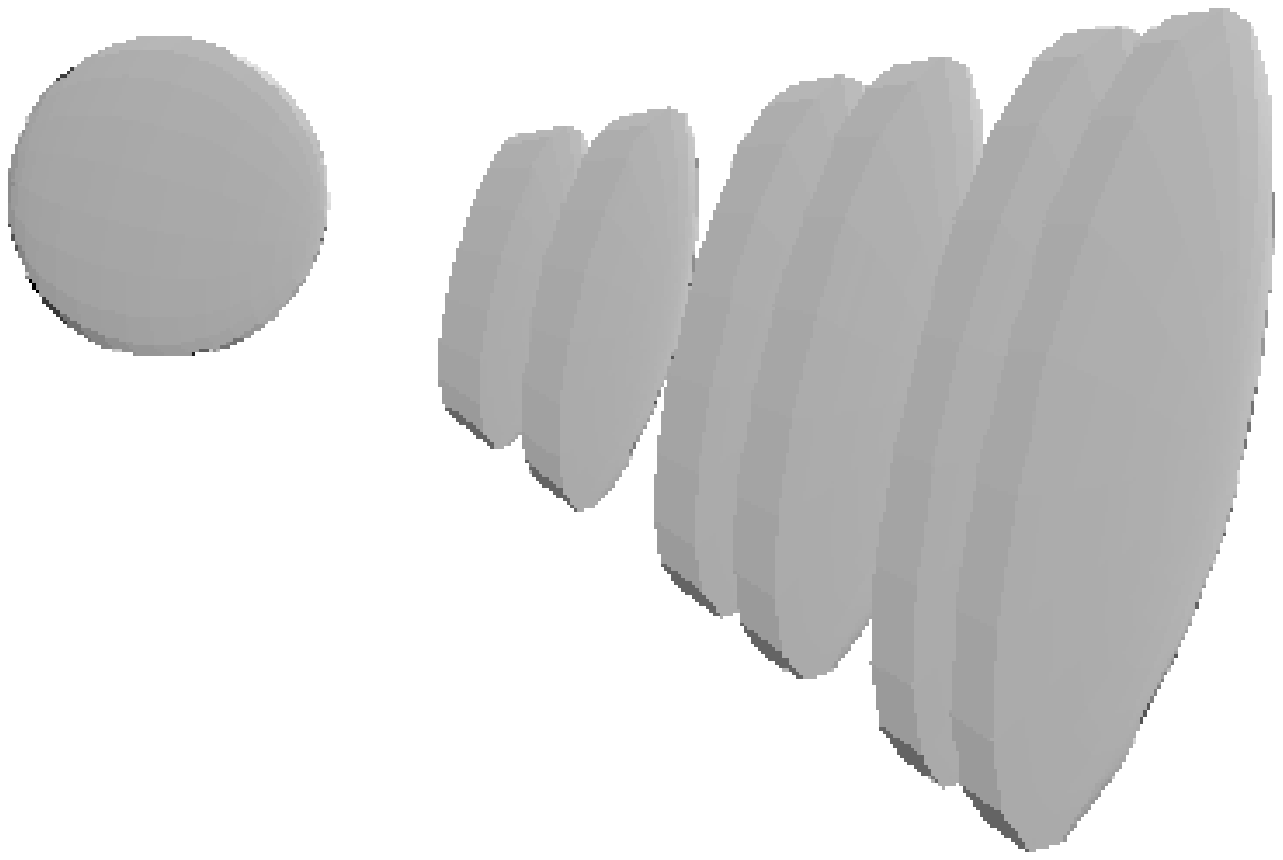}}
\vspace{10pt}
\caption{ Producing variability by external shocks (left) or
internal shocks (right). In the external shocks scenario, the
variability is produced by irregularities in the surrounding.  If the
surrounding consists of a low density medium that contains high
density clouds, then whenever the shell hits one of the clouds a peak
in the emission is produced.  The number of clouds within the
observable cone (of angular size $1/\gamma$ due to relativistic
beaming) should therefore roughly be the number of observed peaks. The
source itself, in this model, needs to produce only a single shell in
a single (simple) explosion. However, the external shocks scenario has
low efficiency, due to the small total surface area of the clouds when
compared to the area of the shell. In the internal-shocks case, the
temporal structure arises from the source, i.e. the source produces a
more complex explosion. There is no efficiency problem, provided that
the relative Lorentz factor between shells is large.
\label{fig:extint}}
\end{figure}

Internal shocks do not suffer from these problems. Detailed
calculations show that the observed temporal structure from
internal shocks closely follows the operation of the inner engine
that generated the shells (Kobayashi, Piran, \& Sari 1997).  In this scenario, the source
must be variable on time scales shorter than a second and last for as
long as $100$ seconds, just as the bursts themselves.

The efficiency of internal shocks is largely determined by the ratio
of Lorentz factors between different shells which are colliding with
each other. The larger the ratio, the larger the efficiency. A simple
scenario that demonstrates this is the case of two equal mass shells
with Lorentz factors $\gamma_1 \gg \gamma_2 \gg 1 $. Conservation of
energy and momentum in a collision between the shells leads to a
Lorentz factor which is the geometric mean of the initial ones
$\sqrt{\gamma_1 \gamma_2}$. Therefore, the energy left in the system
as non-thermal is a small fraction $\sqrt{\gamma_2/\gamma_1}$ of the
initial energy. Beloboradov (2000) has argued that if large
Lorentz factor ratios are allowed, the internal shock efficiency is
only limited by the fraction of energy in the shock given to the
radiating electrons.  Kobayashi and Sari (2001) have then shown
that multiple collisions between shocks may result in `ultra
efficient' internal shocks, in the sense that even more than the
fraction of energy given to electrons can be radiated away.

The mechanism by which the thermal energy produced by internal shocks
is converted to radiation is almost certainly synchrotron and inverse
Compton, since these are the dominant radiation mechanisms at the low
densities involved. While both mechanisms probably take place, it is
actually not very clear which of the two produces the observed
radiation. Synchrotron emission is for several reasons preferred 
(Sari, Narayan \& Piran 1996; Sari \& Piran 1997b) and
inverse Compton probably produces a higher energy component.

\section{The afterglow: theory \label{sec:afttheo}}

After the internal shocks produce the GRB, the shell interacts with
the surrounding medium and decelerates. Again it emits radiation by
synchrotron and inverse Compton. As the flow decelerates, the emission
shifts to lower and lower frequencies. This emission, the afterglow,
may last on detectable levels for years after the GRB itself!

Afterglow was predicted well before it was observed
(Paczy\'nski \& Rhoads 1993; Katz, 1994; Vietri, 1997; M\'esz\'aros \& Rees 1997). The afterglow theory is relatively
simple. It deals with the emission on timescales much longer than that
of the GRB. The details of the complex initial conditions are
therefore forgotten and the condition of the GRB remnant can be
described by a self-similar solution with a small number of
parameters, such as the total energy and the external density. It is
assumed that the electrons are accelerated by the shock into a
power-law distribution of electron Lorentz factors $N(\gamma
_{e})\propto \gamma _{e}^{-p}$ for $\gamma _{e}>\gamma _{m}$. The
lower cutoff $\gamma_{m}$ of this distribution is set by the
assumption that the electrons acquire a fixed fraction, $\epsilon_e$,
of the thermal energy. It is also assumed that a considerable magnetic
field is built behind the shock, which is again characterized by a
certain fraction $\epsilon _{B}$ of equipartition. The energy
density behind a relativistic shock is given by $4 \gamma^2 n m_p
c^2$, where $n=n_1$~cm$^{-3}$ is the proton number density behind the
shock, $\gamma$ is the Lorentz factor of the fluid behind the shock,
and $m_p$ is the proton mass. These equipartition assumptions then
result in
\begin{equation} 
\label{eq:gamma_m}
\gamma_m = {p-2 \over p-1} {m_p \over m_e} \epsilon_e \gamma 
\cong 630 \epsilon_e \gamma
\end{equation}
\begin{equation}
\label{eq:B}
B=0.4 \sqrt{\epsilon_B n_1} \gamma \,\,{\rm Gauss,}
\end{equation}
where $B$ is the magnetic field, and $m_e$ is the electron mass.  The
relativistic electrons then emit synchrotron radiation which produces
the observed afterglow. The broadband spectrum of such afterglow
emission was given by Sari, Piran \& Narayan (1998).

\begin{figure}[tbp]
\centerline{\epsfxsize15pc\epsfbox{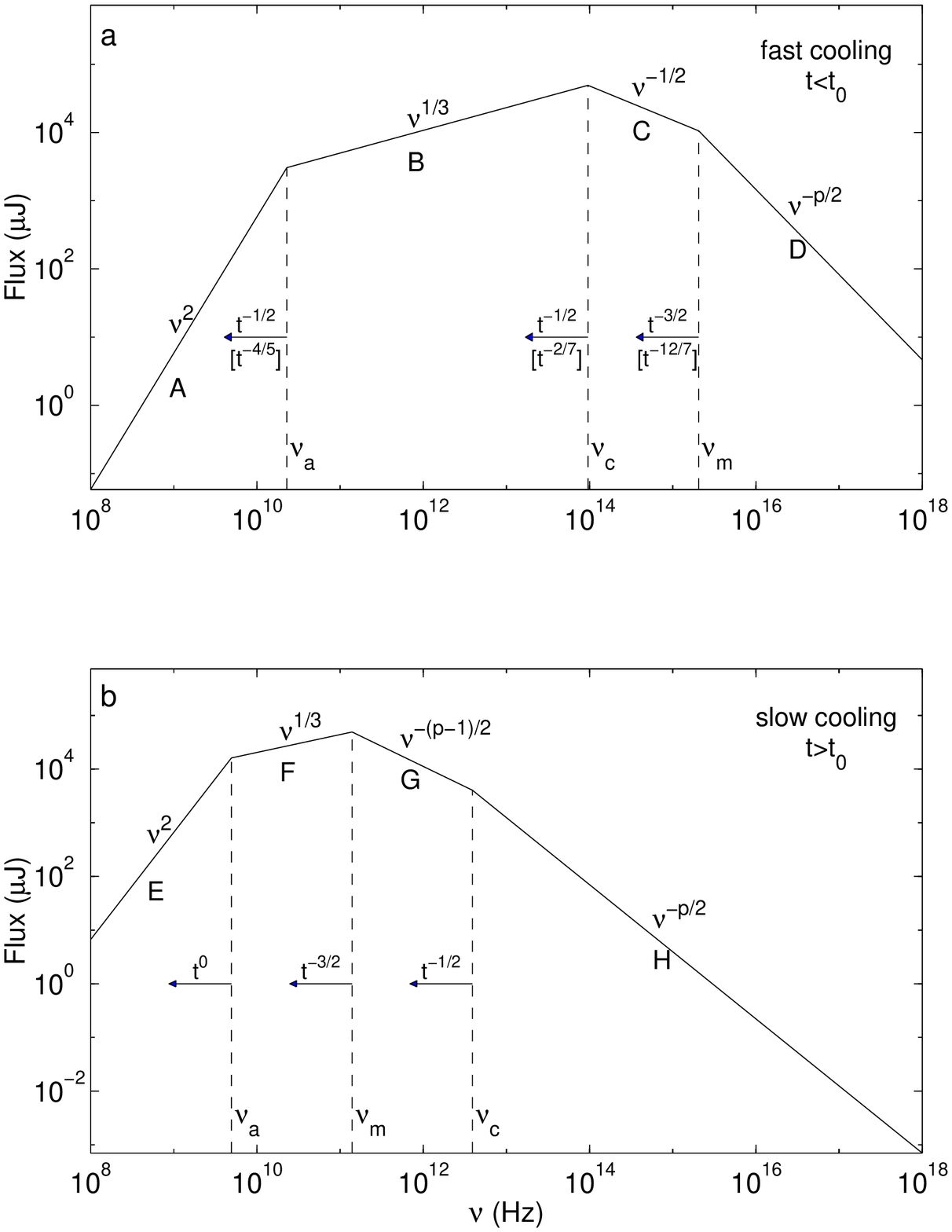}\ \ \epsfxsize14.5pc \epsfbox{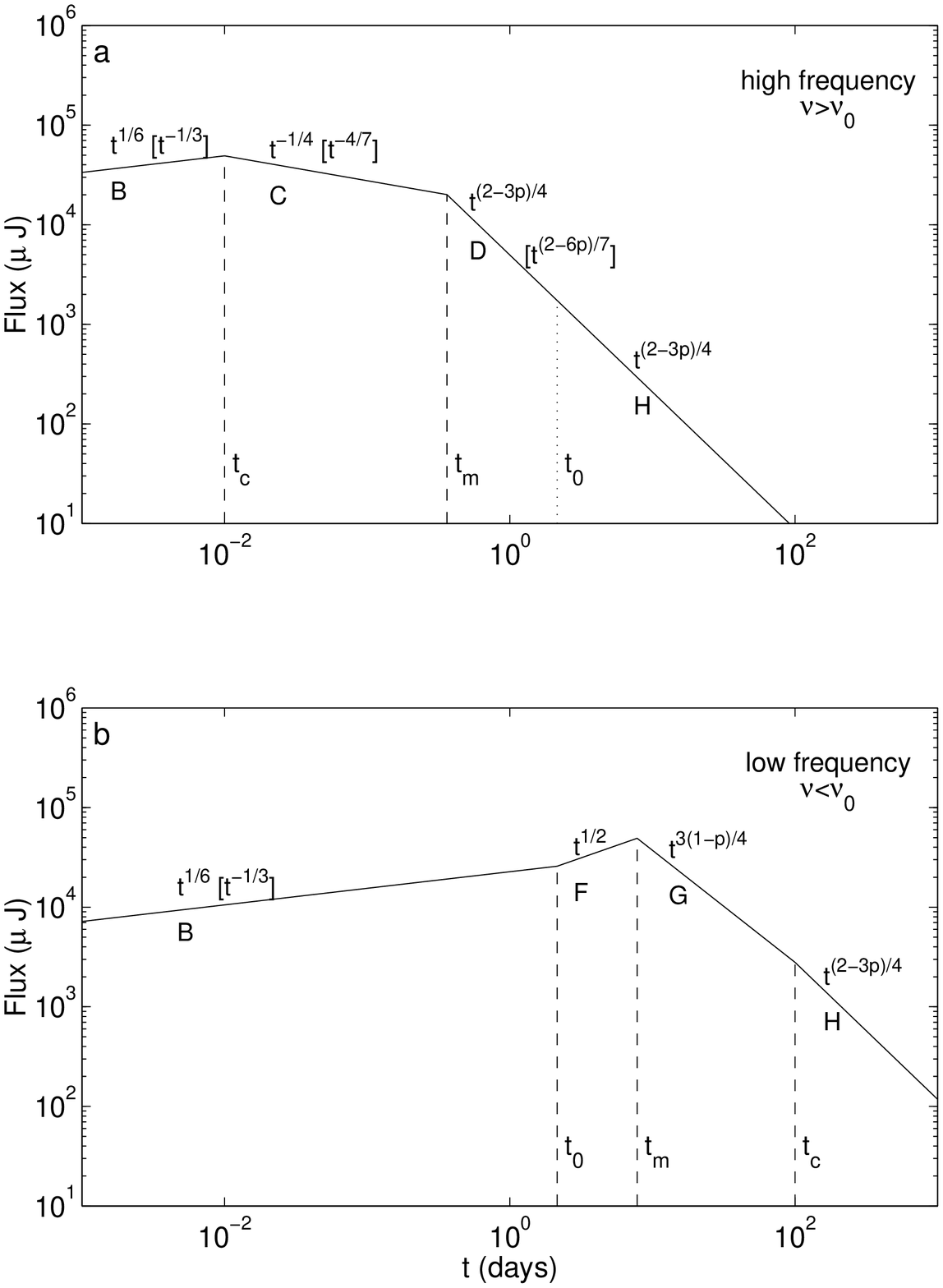}}
\vspace{10pt}
\caption{ Theoretical spectra (left) and light curves (right) of
synchrotron emission from a powerlaw distribution of electrons for the
case of a constant density ambient medium and a spherical explosion.
For most cases $p=2.2-2.5$ fits the observed spectra and
lightcurves well .
\label{fig:aftspec}}
\end{figure}

The afterglow synchrotron spectrum can be fully described by the
electron energy index $p$, the peak flux $F_m$ and three
characteristic frequencies ($\nu_m$, $\nu_c$, $\nu_a$): \\

{\noindent \bf (I)} $\nu_m$ is the synchrotron frequency of the
minimum energy electron, with Lorentz factor $\gamma _{m}$. From
synchrotron theory $\nu_m \cong (e B/2 \pi m_e c)\gamma_m^2$ in the
local frame of the fluid; here $e$ is the electron charge.
Transforming this to the observer frame (blue shifted by the Lorentz
factor and redshifted by a factor of $[1+z]$) and using equations
\ref{eq:gamma_m} and \ref{eq:B} we obtain
\begin{equation}
\nu_m=1.4\times 10^{13}{\rm Hz} \left( 1+z \right)^{-1} \left(
\frac{\epsilon _{e}}{0.1}\right) ^{2}\left( \frac{\epsilon
_{B}}{0.1}\right) ^{1/2}(\frac{\gamma}{10})^{4}n_{1}^{1/2}.
\end{equation}

{\noindent \bf (II)} The cooling time of an electron is inversely
proportional to its Lorentz factor $\gamma _{e}$.  Therefore,
electrons with a Lorentz factor higher than a critical Lorentz factor
$\gamma _{e}>\gamma _{c}$ cool on the dynamical timescale of the
system. This characteristic Lorentz factor is given by the condition
$\sigma _{T}c\gamma ^{2}\gamma _{c}^{2}B^{2}t\gamma/6\pi(1+z) =\gamma
_{c}m_{e}c^{2}$, and corresponds to the `cooling frequency'
\begin{equation}
\nu _{c}=1.2\times 10^{13}{\rm Hz} \left( 1+z \right) \left(
\frac{\epsilon _{B}}{0.1}\right) ^{-3/2}\left( \frac{\gamma
}{10}\right) ^{-4}n_{1}^{-3/2}t_{days}^{-2},
\label{nucf}
\end{equation}
where $t_{days}$ is the observer time in days. Here we have also
taken into account that time is redshifted.

{\noindent \bf (III)} Below some critical frequency $\nu _{a}$ the
flux is self-absorbed and is given by the Rayleigh-Jeans portion of a
blackbody spectrum\footnote{ Granot, Piran \& Sari (2000b) have
found that if $\nu_{c} < \nu_{m}$, then the self-absorption
frequency actually splits into two: $\nu_{ac}$ and $\nu_{sa}$, where
an optical depth of unity is produced by non-cooled electrons and all
electrons, respectively. In between these two frequencies the spectral
slope is $\nu^{11/8}$.}. The self-absorption frequency is given by
\begin{equation}
\label{nusa1}
\nu _{sa}=93\,{\rm GHz} \left( 1+z \right)^{-13/5} \left(
\frac{\epsilon _{B}}{0.1}\right) ^{6/5}\left( \frac{\gamma
}{10}\right) ^{28/5}n_{1}^{9/5}t_{days}^{8/5},
\end{equation}
if $\nu_c<\nu_m$, and by 
\begin{equation}
\label{nusa2}
\nu _{sa}=87\,{\rm GHz} \left( 1+z \right)^{-8/5} \left(
\frac{\epsilon _{e}}{0.1}\right) ^{-1} \left( \frac{\epsilon
_{B}}{0.1}\right) ^{1/5} \left( \frac{\gamma }{10}\right)
^{8/5}n_{1}^{4/5}t_{days}^{3/5},
\end{equation}
if $\nu_c>\nu_m$. \\

{\noindent \bf (IV)} The normalization of the spectrum is given by the
total number of radiating electrons $4\pi R^3 n_1/3$ times the peak
flux from a single electron, resulting in
\begin{equation}
F_{m}=220\,{\rm mJy} (1+z)^{-2}  d_{L,28}^{-2}\left( \frac{\epsilon _{B}}{0.1}%
\right) ^{1/2}\left( \frac{\gamma }{10}\right) ^{8}n_{1}^{3/2}t_{days}^{3},
\end{equation}
where $d_{L,28}$ is the luminosity distance in units of 10$^{28}$cm.

The broadband spectrum of the well studied GRB\,970508 (Galama et al. 1998b)
is in very good agreement with the theoretical picture. Note that the
derivation above is quite general. It does not depend either on the
surrounding density profile or on the geometry of the event. Both
these effects are hidden in the evolution of the fluid Lorentz factor
$\gamma$, and the particle density $n_1$ as a function of time.

The evolution of this spectrum as a function of time depends on the
hydrodynamics. The simplest, which describes the observations in some
cases quite well, is the adiabatic model with a constant density
surrounding medium. The rest mass collected by the shock at radius $R$
is about $R^{3}\rho$, where $\rho$ is the mass density. On average,
the particles move with a Lorentz factor of $\gamma^{2}$ in the
observer frame (one factor of $\gamma$ is the bulk motion and the
other is the random thermal motion). Therefore, the total energy is
given by $E\propto \gamma ^{2}R^{3}\rho c^{2}$. Assuming that the
radiated energy is negligible compared to the energy of the flow, we
obtain that $\gamma \propto R^{-3/2}$ or in terms of the observer
time, $t=R/\gamma^2c$, we get $\gamma \propto t^{-3/8}$.

$$\nu_m=6 \times 10^{15}\, {\rm Hz \ } (1+z)^{1/2}E_{52}^{1/2}\epsilon_{e}^2
\epsilon_B^{1/2}t_{\rm days}^{-3/2}$$

$$\nu_c=9 \times 10^{12}\, {\rm Hz\ } (1+z)^{-1/2}
\epsilon_B^{-3/2}n_1^{-1} E_{52}^{-1/2}t_{\rm days}^{-1/2}$$

$$\nu_{sa}=2 \times 10^{9}\,{\rm Hz\ } (1+z)^{-1}\epsilon_{e}^{\;
-1}\epsilon_B^{1/5} n_1^{3/5} E_{52}^{1/5}$$

$$F_m=20\,{\rm mJy\ } (1+z) \epsilon_B^{1/2} n_1^{1/2} E_{52}
d_{L28}^{-2}$$

If, on the other hand (Chevalier \& Li 1999), the density drops as $R^{-2}$ (as is expected
if the surrounding is a wind produced earlier by the progenitor of the
burst) we get $\gamma \sim t^{-1/4}$. Choosing the parameter $A_{*}$
to define the normalization of the density as $\rho R^2=A_{*} 5 \times
10^{11} A_{*}$ gr/cm results in

$$\nu_m=1.7 \times 10^{14}\, {\rm Hz\ } (1+z)^{1/2}E_{52}^{1/2}\epsilon_{e}^2
\epsilon_B^{1/2} t_{\rm days}^{-3/2}$$

$$\nu_c=7 \times  10^{11}\, {\rm Hz\ } (1+z)^{-3/2}
\epsilon_B^{-3/2}A_{*}^{-2} E_{52}^{1/2}t_{\rm days}^{1/2}$$

$$\nu_{sa}=1.5  \times 10^{10}\, {\rm Hz\ }
(1+z)^{-2/5}\epsilon_{e}^{-1}\epsilon_B^{1/5} A_{*}^{6/5}
E_{52}^{-2/5}t_{\rm days}^{-3/5}$$

$$F_m=180\, {\rm mJy\ } (1+z)^{3/2} \epsilon_B^{1/2}
A_{*}E_{52}^{1/2}t_{\rm days}^{-1/2}d_{L28}^{-2}$$

These simple scalings, for the case of a constant density ambient
medium, lead to the spectral evolution given in Figure
\ref{fig:aftspec}. The derivations above use a very simple description
of the flow. It represents the fluid as if it had a single magnetic
field strength and a single Lorentz factor $\gamma$ and all of the
material is moving directly towards the observer. Also, a very
approximate description of the synchrotron emission was used. In
reality, of course, the situation is more complicated. There are two
effects that must be taken into account. The most dramatic one is the
fact that matter slightly off the line of sight does not move directly
towards the observer (Waxman 1997b; Panaitescu \& M\'esz\'aros 1998; Sari 1998). The amount of Lorentz boost
from that matter is reduced. Secondly, fluid elements at different
distances from the shock have somewhat different Lorentz factors,
magnetic fields and electron energies. These variations can be
estimated using the self-similar solution of Blandford and McKee
(1976). The outcome of these more detailed calculations are the
same scaling laws, but with a more accurate coefficient for the break
frequencies as well as an estimate of the shape of the spectrum around
each break frequency (Granot, Piran \& Sari 2000; Gruzinov \& Waxman 1999; Granot, Piran \& Sari 1999). The equations given
above already take these effects into account, and the coefficients
given are accurate for $p=2.2$.

The above scalings assumed adiabatic evolution. At first sight one might
think that if the fraction of energy given to the electrons,
$\epsilon_e$, is less than unity, then perhaps only a small fraction
of the energy can be radiated away. However, the same fireball energy
is given again and again to newly shocked electrons. Each time, a
fraction $\epsilon_e$ can be radiated away, and the overall effect can
be large, much above the fraction $\epsilon_e$.  Energy losses during
the cooling phase can be taken into account (Sari 1998; Cohen, Piran \& Sari 1998) using
$dE/dR= -(16 \pi/3) R^2 \epsilon_e \gamma^2 m_p c^2 n$. This results
in $E=E_0 \times (t/t_0)^{-17\epsilon_e/12}$ for a constant density
environment and $E=E_0 \times (t/t_0)^{-3\epsilon_e/2}$ for a wind
environment. These effects are not taken into account in many models
but may actually have a significant impact if $\epsilon_e$ is not too
far below unity. In the case of GRB\,000926, energy losses appear to
have reduced the energy of the system by a factor of 5 (Harrison et al. 2001).

Given the above hydrodynamic evolution, one can construct light curves
at any given frequency. These will also consist of power laws,
changing from one power law to another once the break frequencies
pass through the observed band. These predicted power law lightcurves
and spectra are in fair agreement with afterglow observations (see
Section \ref{sec:revo}).

We have so far considered synchrotron radiation only. Since the
optical depth of the system is small, most of the synchrotron photons
emitted can be observed. Still, inverse Compton can affect the system
in two ways. First, it may add an observable high-energy
component. This requires a moderately high density.  Second, it may
provide an important cooling mechanism, and alter the synchrotron
spectrum by its effect on $\nu_c$.  The ratio of the inverse Compton
(IC) to synchrotron luminosity (a measure of their relative importance
for cooling) can be computed very generally (Sari, Narayan, \& Piran 1996), in a way
that does not deal with the details of the spectrum, but depends only
on the underlying physical properties of the expanding shock wave.  We
generalize the derivation given by Sari, Narayan, \& Piran (1996) to describe both fast
and slow cooling regimes by introducing a parameter $\eta$, equal to
the fraction of the electron energy that was radiated away (via both
synchrotron and IC emission) (Sari \& Esin 2001).  Then the ratio of
luminosities, in the limit of single scattering, is given by
\begin{equation}
\label{x}
x\equiv \frac{L_{IC}}{L_{syn}} = \frac{U_{rad}}{U_{B}} =
\frac{U_{syn}}{U_{B}} = \frac{\eta U_{e}/(1+x)}{U_{B}} =
\frac{\eta \epsilon_e}{\epsilon_B (1+x)},
\label{ICtoSYN}
\end{equation}
where $U_{syn}$, $U_B$ and $U_e$ are the energy density of synchrotron
radiation, magnetic field and relativistic electrons, respectively.
Note that in general $U_{syn} = \eta \beta U_e/(1+x)$, where $\beta$
is the velocity of material behind the shock front (in the frame of
the shock); however, for a relativistic shock $\beta\cong 1$. The
importance of inverse Compton therefore diminishes quickly when the
fireball becomes non-relativistic.
 
Solving Eq. (\ref{x}) for $x$ we obtain
\begin{equation}
x=\frac {-1+\sqrt{1+4 \frac {\eta\epsilon_e} {\epsilon_B} }} {2}
\cong
\left\{ \begin{array} {ll}
\frac{\eta \epsilon_e}{\epsilon_B}, & \mathrm{if \quad}
\frac{\eta \epsilon_e}{\epsilon_B} \ll 1, \\
\left(\frac{\eta \epsilon_e}{\epsilon_B}\right)^{1/2}, & \mathrm{if \quad}
\frac{\eta \epsilon_e}{\epsilon_B} \gg 1.
\end{array} \right.
\label{approxx}
\end{equation}

Modeling afterglow data often suggests that $\epsilon_e \gg
\epsilon_B$ and therefore inverse Compton may be of importance.

\section{The afterglow revolution \label{sec:revo}}

Motivated by the prediction of a late-time softer radiation (the
afterglow), several groups executed rapid radio follow-up observations
of GRB error boxes. Detection of a radio afterglow seemed most
promising. Not only does the large field of view match well with the
large error boxes (several degrees) that were then available on short
time scales (within a day), but maximum light was also expected to
occur later at longer wavelengths. The best (pre-BeppoSAX era) limits
on such afterglow radio emission were obtained for GRB\,940301. This
GRB triggered an extensive multi-wavelength campaign with ground based
optical and radio observatories from the BATSE / COMPTEL / NMSU Rapid
Response Network (McNamara et al. 1995). No obvious candidate radio
counterparts were found (Frail et al. 1994; Koranyi et al. 1995; Galama et al. 1997a).

\subsection{The first identifications }

The breakthrough came in early 1997, when the Wide-Field Cameras
(WFCs; Jager et al. 1993) onboard the Italian-Dutch satellite BeppoSAX
(Piro, Scarsi \& Butler 1995) obtained their first quickly available (within hours)
accurate positions of GRBs (several arcminutes). This allowed rapid
follow-up observations which led to the discoveries of X-ray
(Costa et al. 1997), optical (van Paradijs et al. 1997), millimeter (Bremer et al. 1998) and
radio (Frail et al. 1997) counterparts of GRBs. These observations quickly
settled the distance controversy. The first transient optical
counterpart, of GRB\,970228, is in a faint galaxy with $\sim
0.8^{\prime\prime}$ diameter (Sahu et al. 1997).  And, detection of
absorption features in the OT's spectrum of GRB\,970508 (Metzger et al. 1997)
established that this event was at a redshift greater than $z =
0.835$. GRBs come from `cosmological' distances and are thus extremely
powerful events. They are by far the most luminous photon sources in
the Universe, with (isotropic equivalent) peak luminosities in
$\gamma$ rays up to $10^{52}$ erg/s, and total energy budgets up to
several $10^{53-54}$ erg (Kulkarni et al. 1998; Kulkarni et al. 1999a) (but see
Section \ref{sec:beatheo} and \ref{sec:beaobs} for a discussion of
collimated outflow, which reduces the inferred total energy).  Within
the first day, the optical emission is usually brighter than 20th
magnitude (some 10 mag brighter [absolute] than the brightest
supernovae) and therefore small telescopes can play an important role
in measuring the lightcurve. Today, a large worldwide collaboration
is observing these events and the data are submitted to the
Gamma-Ray Burst Coordinates Network in near-real time, allowing other observatories
to react rapidly.

\subsection{Confirmation of the relativistic blast-wave model}

A stringent test of the relativistic blast wave model came with the
discovery of X-ray (Costa et al. 1997) and optical afterglow following
GRB\,970228 (van Paradijs et al. 1997; Galama et al. 1997b). The X-ray and optical afterglows of
GRB\,970228 show a power-law temporal decay; this is a trend observed
in all subsequent X-ray and optical afterglows, with power-law
exponents in the range 1 to 2.  

Let us first concentrate on the forward shock and assume slow cooling
(the bulk of the electrons do not radiate a significant fraction of
their own energy and the evolution is adiabatic); this appears
applicable to some observed GRB afterglows at late times ($t >$ 1 hr).
The simplest assumption is that of spherical symmetry and a constant
ambient density.  As both the afterglow's spectrum and the temporal
evolution of the break frequencies $\nu_{a}$, $\nu_{m}$,
$\nu_{c}$ are, in the relativistic blast wave model, power laws
(see Section \ref{sec:afttheo}), the evolution of the flux is also a
power law in time. For example, for $\nu_{m} \le \nu \le
\nu_{c}$, the decay of the flux is {\scriptsize $F_{\nu} \propto
t_{\rm obs}^{-3(p-1)/4}$}, and the power law spectral slope $\alpha$
relates to the spectral slope $\beta$ as $\alpha = -3/2
\beta$. Several authors (Wijers, Rees, \& {M\'esz\'aros}, 1997; Reichart 1997; Waxman 1997a) showed that to first
order this model describes the X-ray and optical afterglow of
GRB\,970228 very well.


GRB\,970508 was the first GRB with a radio counterpart
(Frail et al. 1997). The radio light curves (8.5 and 4.9 GHz) show large
variations on time scales of less than a day, but these damp out after
one month. This finds a viable explanation in interstellar
scintillation (stochastic refraction and diffration by the fluctuations in
the interstellar medium electron density
between the source and the observer). The damping of the fluctuations
can then be understood as the effect of source expansion on the
diffractive interstellar scintillation. Thus a source size of roughly
10$^{17}$ cm was derived (at 3 weeks), corresponding to a mildly
relativistic expansion of the shell (Frail et al. 1997).

GRB\,970508 remains one of the best observed afterglows: the radio
afterglow was visible at least 368 days (and with 2.5 $\sigma$ significance on day 408.6
(Frail, Waxman \& Kulkarni 2000)), and the optical afterglow up to $\sim$ 450 days
(e.g. Fruchter et al. 2000; Galama et al. 1998a; Castro-Tirado et al. 1998). In addition millimeter
(Bremer et al. 1998), infrared and X-ray (Piro et al. 1998) counterparts were
detected. 
These multiwavelength
observations allowed the reconstruction of the broad radio to X-ray
spectrum for this GRB (Galama et al. 1998b) 
Galama et. al. (1998b) found that the `standard' model provides
a successful and consistent description of the afterglow observations
over nine decades in frequency, ranging in time from the event until
several months later. The synchrotron afterglow spectrum of this GRB
allows measurement of the electron energy spectrum $p$, the three
break frequencies ($\nu_{a}$, $\nu_{m}$ and $\nu_{c}$), and the flux
at the peak, $F_{m}$. For GRB\,970508 the redshift $z$ is also
known, and all blast wave parameters could be deduced: the total
energy (per unit solid angle) E = 3.5$\times10^{52}$ erg, the ambient
(nucleon) density $n_1 \approx 5$, the fraction of the energy in
electrons $\epsilon_{e} \approx 0.5$ and that of the magnetic field
$\epsilon_B = 0.01$ (Wijers \& Galama 1999; Granot, Piran, \& Sari 1999). The numbers themselves are
uncertain by an order of magnitude, but the result shows that the
`standard' model fits the expectations very well.

Following these first attempts at modeling the broad-band afterglow
more detailed modeling efforts have been made. For example, Panaitescu
and Kumar (2001a) have modeled a sample of GRBs with relativistic
jets (see Section \ref{sec:beatheo} and \ref{sec:beaobs} for a detailed
discussion on jets) and find: typical energies of $10^{50}-10^{51}$
erg, ambient densities ranging from $10^{-3}-10$ cm$^{-3}$, beaming
angles ranging between $1^{\circ} - 4^{\circ}$, and that a wind-like
ambient medium can in some cases be ruled out  GRB\,000301C was
modeled with a hard electron-energy distribution (Panaitescu 2001); $p$ =
1.5) (but see Berger et al. 2000) and GRB\,010222 also requires a hard
electron energy distribution (Galama et al. 2001). Evidence has been
presented for an inverse Compton emission component in the afterglow
of GRB\,000926 (Harrison et al. 2001).


The highly relativistic nature of the GRB source (Galama et al. 1999) can
once more be seen in the extreme brightness temperature of the
GRB\,990123 optical flash (Akerlof et al. 1999) $T_b\ge 10^{17}$\,K; see
Sect \ref{sec:revobs}) which by far exceeds the Compton limit of
$10^{12}$\,K. In this case the optical signal from GRB\,990123 was
some 18 mag brighter (absolute) than the brightest supernovae. The
extreme brightness can be explained by emission from the reverse shock
(see Section \ref{sec:revtheo}).

\section{Collimated outflow (jets): theory \label{sec:beatheo}} 

The hydrodynamic evolution described in Section \ref{sec:afttheo},
assumed spherical symmetry. However, many astrophysical phenomena,
especially those involving extreme energetics, are not spherical but in
the form of jets.  As we will see, this is most probably the case also
for GRBs.

Jets have been discussed extensively in the context of GRBs. First,
the similarity between some of the observed features of blazars and
AGNs led to the speculation that jets also appear in GRBs
(Paczy\'nski 1993). Second, the regions emitting the GRBs as well as the
afterglow must be moving relativistically. The emitted radiation is
strongly beamed, and we can observe only a region with an opening
angle 1$/\gamma$ off the line of sight. Emission outside of this very
narrow cone is not observed. These considerations have led to
numerous speculations on the existence of jets and to attempts to
search for the observational signature of jets both during the GRB
phase (Mao \& Yi 1994) and in the context of the afterglow
(Rhoads 1997; Rhoads 1999; M\'esz\'aros, Rees \& Wijers 1998). 

We begin by clarifying some of the confusing terminology. There are
two distinct but related effects.  The first, {\it `jets'},
describes scenarios in which the relativistic flow emitted from the
source is not isotropic but collimated into a finite solid
angle. The term jet refers to the geometrical shape of the
relativistic flow emitted from the inner engine. The second effect is
that of {\it `relativistic beaming'}. The radiation from any object
that is radiating isotropically in its own rest frame, but moving with
a large Lorentz factor $\gamma$ in the observer frame, is beamed into
a small angle $1/\gamma$ around its direction of motion. This is an
effect of special relativity. It has nothing to do with the ejecta's
geometry (spherical or jet) but only with the fact that the ejecta is
moving relativistically.  The effect of relativistic beaming allows an
observer to see only a small angular extent, of size $1/\gamma$
centered around the line of sight. Since we know the flow is
ultra-relativistic (initially $\gamma>100$), there is no question that
the relativistic beaming effect is always relevant for GRBs. The
question we are interested in is that of the existence of `jets'.

The idealized description of a jet is a flow that occupies only a
conical volume with half opening angle $\theta_0$. In fact, the
relativistic dynamics is such that the width of the material in the
direction of its propagation is much smaller than its distance from
the source by a factor of $1/\gamma^2$. The flow, therefore, does not
fill the whole cone. Instead it occupies only a thin disk at its base,
looking more like a flying pancake (Piran 1999) (see Figure
\ref{fig:extint}).  If the `inner engine' emits two such jets in
opposite directions then the total solid angle towards which the flow
is emitted is $\Omega=2\pi \theta_0^2$. Whether the relativistic flow
is in the form of a jet or a sphere has three important
implications. \\

\noindent {\bf The Total Emitted Energy.} Optical observations of
afterglows enabled redshift determinations, and therefore reasonably
accurate estimates of the distance, $D$, to these events (the
uncertainty is now in the cosmological parameters of the
Universe). The so called `isotropic energy' can then be inferred from
the fluence $F$ (the total observed energy per unit area at earth) as
$E_{iso}=4\pi D^2 F$ (taking cosmological corrections into account,
$D=d_L/\sqrt{1+z}$ where $d_L$ is the luminosity distance and $z$ is
the redshift). The numbers obtained in this way range from $10^{51}$
erg to $10^{54}$ erg with the record of $3\times 10^{54}$ erg held by
the famous GRB\,990123. These huge numbers approach the equivalent
energy of a solar mass, all emitted in a few tens of seconds!

These calculations assumed that the source emitted the
same amount of energy in all directions. If instead the emission
is confined to some solid angle $\Omega$ then the true energy is
$E=\Omega D^2 F$. As we show later $\Omega$ is very weakly constrained
by the GRB itself and can be as low as $10^{-6}$. If so the true
energy in each burst $E \ll E_{iso}$. We will show later that
interpretation of the multi-wavelength afterglow lightcurves indeed indicates
that some bursts are jets with solid angles considerably less than $4 \pi$.
The isotropic energy estimates may be fooling us by a few orders of
magnitudes! Clearly this is of fundamental importance when considering
models for the sources of GRBs. \\

\noindent {\bf The Event Rate.} In its glory days, BATSE detected
about one burst per day. With the help of several redshift
measurements, or alternatively, with the use of the cumulative
brightness distribution (the Log N-Log S curve), this translates to
about $10^{-7}$ bursts per year per galaxy or $0.5$
bursts/Gpc$^{-3}$/year (Schmidt 1999; Schmidt 2001). However, if the emission is
collimated to $\Omega \ll 4\pi$ then we do not see most of the
events. The true event rate is then larger than that measured by BATSE
by a factor of $4\pi /\Omega$. Again this is of fundamental
importance.  Clearly, the corrected GRB event rate must not exceed
that of compact binary mergers or the birth rate of massive stars if
these are to produce the majority of the observed GRBs 

\noindent {\bf The Physical Ejection Mechanism.} Different physical
models are needed to explain collimated and isotropic emission.  For
example, in the collapsar model (e.g. MacFadyen \& Woosley 1999b), relativistic
ejecta that are believed to create the GRB are produced only around
the rotation axis of the collapsing star with a half opening angle of
about $\theta_0 \cong 0.1$. Such models would have difficulties
explaining isotropic bursts as well as very narrow jets.

\subsection{The jet-break}
As the afterglow evolves, $\gamma$ decreases and it will eventually
fall below the initial inverse opening angle of the jet. The observer
will notice that some of the sphere is missing from the fact that less
radiation is observed. This effect alone will produce a significant
break, steepening the lightcurve decay by a factor of $\gamma^2
\propto t^{-3/4}$ even if the dynamics of each fluid element has not
changed. The transition should occur at the time $t_{jet}$ when
$1/\gamma \cong \theta_0$. Observing this time can therefore provide
an estimate of the jet's opening angle according to
\begin{equation}  \label{t_jet}
t_{{\rm jet}}\approx {6.2{\rm hr} (1+z)} (E_{52}/n_{1})^{1/3}(\theta _{0}/0.1)^{8/3}
.
\end{equation}

Additionally, Rhoads (1999) has shown that at about the same time
(see however Panaitescu \& M\'esz\'aros 1999; M\'esz\'aros \& Rees 1999; Moderski, Sikora \& Bulik 2000), the jet will begin to spread
laterally so that its opening angle $\theta (t\grave{)}\sim 1/\gamma$.
The ejecta now encounter more surrounding matter and decelerate
faster than in the spherical case. The Lorentz factor then decays
exponentially with the radius and as $\gamma\propto t^{-1/2}$ with
observed time. Taking this into account, the observed break is even
more significant. The slow cooling spectrum given in Figure
\ref{fig:aftspec} evolves with decreasing peak flux $F_{m} \propto
t^{-1}$ and the break frequencies evolve as $\nu_{m} \propto t^{-2}$,
$\nu_{c} \propto t^0$ and $\nu_{a} \propto t^{-1/5}$.  This translates
to a temporal decay at a given frequency given in Table
\ref{t:afterglow}.

\begin{table}[!ht]
\begin{center}
\begin{tabular}{|c||c||c|c|}
\hline
& spectral index & \multicolumn{2}{|c|}{light curve index $\alpha$, $F_{\nu}\propto
t^{-\alpha}$} \\
& $\beta$, $F_{\nu}\propto \nu^{-\beta}$ & sphere & jet \\ \hline\hline
$\nu<\nu_{a}$ & $\beta=-2$ & $\alpha=-1/2$ & $\alpha=0$ \\ \hline
$\nu_{ a}<\nu<\nu_{ m}$ & $\beta=-1/3$ & $\alpha=-1/2$ & $\alpha=1/3$ \\ \hline
&  & $\alpha=3(p-1)/4\cong 1.05$ & $\alpha=p\cong 2.4$ \\
\raisebox{1.5ex}[0pt]{$\nu_{ m}<\nu<\nu_{ c}$} & \raisebox{1.5ex}[0pt]{$(p-1)/2
\cong0.7$} & $\alpha=3\beta/2$ & $\alpha=2\beta+1$ \\ \hline
&  & $\alpha=(3p-2)/4\cong 1.3$ & $\alpha=p\cong 2.4$ \\
\raisebox{1.5ex}[0pt]{$\nu>\nu_{ c}$} & \raisebox{1.5ex}[0pt]{$ p/2
\cong1.2$} & $\alpha=3\beta/2-1/2$ & $\alpha=2\beta$ \\ \hline
\end{tabular}
\end{center}
\caption{ The spectral index $\beta$ and the temporal index $\alpha$ as
functions of $p$ for a spherical and a jet-like evolution. Typical values are
quoted using $p=2.4$. The parameter-free relation between $\alpha$ and
$\beta$ is given for each case (eliminating $p$). The difference in $\alpha$
between a jet and a sphere is always substantial at all frequencies.}
\label{t:afterglow}
\end{table}

The jet break is a hydrodynamic one. It should therefore appear at the
same time at all frequencies - an achromatic break\footnote{Sari 1997
(Sari 1997), argued that there may be about a factor of two difference
in the effective transition time between the four different spectral
regimes (e.g. below or above $\nu_{ m}$) due to the fact that the
emission in these different regimes weighs contributions
from various emission radii.differently }. Though an achromatic break is
considered to be a strong signature of a jet, one should keep in mind
that any other hydrodynamic transition will also produce an achromatic
break. To name a few: the transition from relativistic to
non-relativistic dynamics, a jump in the ambient density or the supply
of new energy from slower shells that catch up with the decelerated
flow. However, the breaks produced by the transition from a spherical-
like evolution (when $1/\gamma<\theta_0$) to a spreading jet have a
well defined prediction for the change in the temporal decay
indices. The amount of break depends on the spectral regime that is
observed. It can be seen from Table \ref{t:afterglow} that the break
is substantial ($\Delta \alpha >0.5$ in all regimes) and should be
easily identified.

\section{Observational evidence for collimated outflow (jets)
\label{sec:beaobs}}

The theory of jet evolution and of the resulting light curves was
worked out before evidence for jets was obtained. In fact, Rhoads
1999), has used this theory to constrain the amount of
collimation in GRB\,970508, which did not show any significant
steepening of the afterglow lightcurve. He concluded that the opening
angle of a jet, if it exists, must be more than 30 degrees.  We note
that if the jet's opening angle is of order unity, the total energy
may still be about an order of magnitude lower than the isotropic
estimate. However, in this case the break will be `hidden' as it will
overlap the transition to non-relativistic dynamics. Based on late
time radio data, it was suggested that this is the case for
GRB\,970508 (Frail, Waxman \& Kulkarni 2000).

The first claim for narrow jets in GRBs came from Sari, Piran and
Halpern (1999). They noted that the observed decays in GRB
afterglows that do not show a break either have a shallow slope
$F_{\nu} \propto t^{-1.2}$ or a very steep slope $F_{\nu} \propto
t^{-2}$. They argued that the rapidly decaying bursts are those in
which the ejecta was a narrow jet and the break in the light curve was
before the first observations. Interestingly, evidence for jets is
found when the inferred energy (without taking jets into account) is
the largest. This implies that the jets account for a considerable
fraction of the wide luminosity distribution seen in GRBs, and the
true energy distribution is less wide than it seems to be.

The predicted light-curve transition (from a regular to a fast decay
caused by a jet) has been observed in the optical afterglow of
GRB\,990123 (Kulkarni et al. 1999a; Castro-Tirado et al. 1999; Fruchter et al. 1999).  However, no evidence for
such an increase in the decay rate was found in near-infrared K-band
observations (Kulkarni et al. 1999a). A similar transition was better sampled in
afterglow data of GRB\,990510; optical observations of GRB\,990510,
show a clear steepening of the rate of decay of the light
simultaneously in all optical bands between $\sim$ 3 hours and several
days (Harrison et al. 1999; Stanek et al. 1999) to roughly F$_{\nu}(t) \propto
t^{-2.2}$. Together with radio observations, which also reveal a
transition, it is found that the transition is very much
frequency-independent; this virtually excludes explanations in terms
of the passage of the cooling frequency, but is what is expected in
case of beaming (Harrison et al. 1999). Harrison et al.(1999) derive a
jet opening angle (from the jet-break time) of $\theta_0 \cong 0.08$,
which for this burst would reduce the total energy in $\gamma$ rays to
$\sim 10^{51}$ erg.


\begin{figure}
\centerline{
\epsfxsize15pc \epsfbox{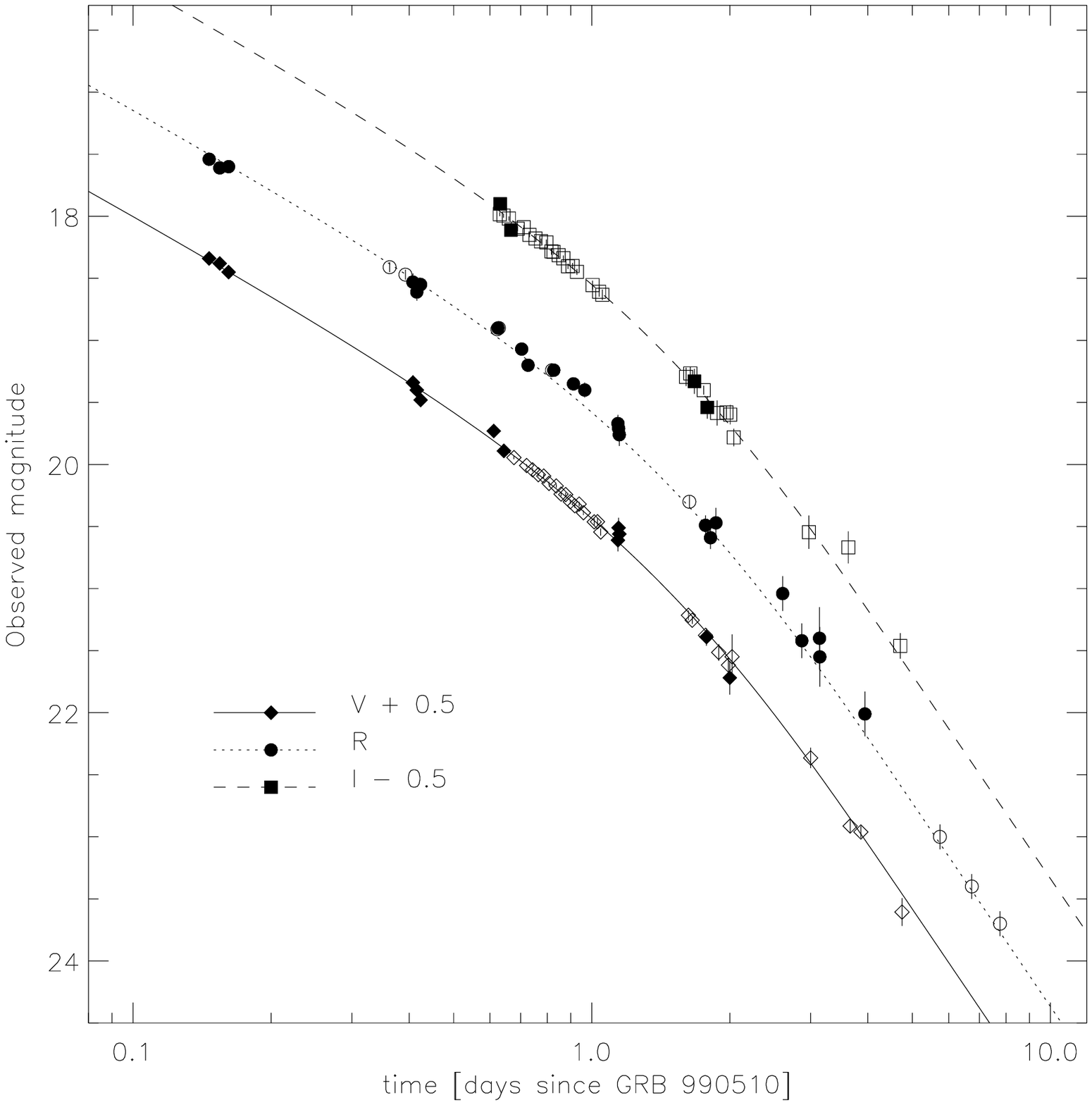}\ \ 
\epsfxsize15pc \epsfbox{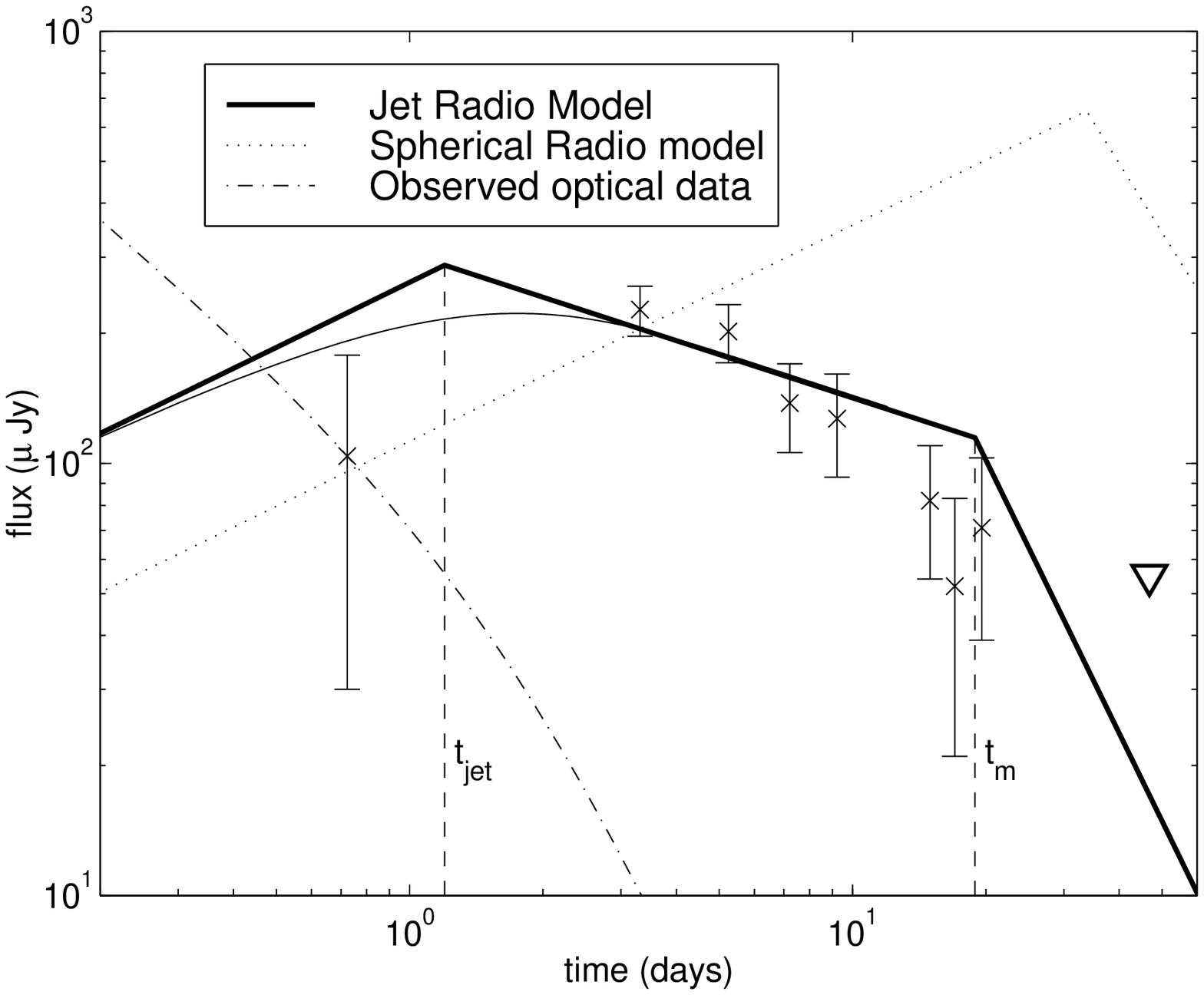}}
\vspace{10pt}
\caption{
\it GRB\,990510, the `classical' case for a `jet': an
achromatic break in optical and radio at $t_{\rm jet} = 1.2$ days
implying a jet opening angle $\theta_0 = 0.08$. The temporal slope
before and after the break agree well with the theory if $p =
2.2$. For this burst the isotropic gamma-ray energy $E_{\rm
iso}=2.9\times 10^{53}$ erg but the `true' total energy is only
$E=10^{51}$ erg. From Harrison et al. (1999). 
}
\end{figure}

\begin{figure}
\centerline{
\epsfxsize20pc \epsfbox{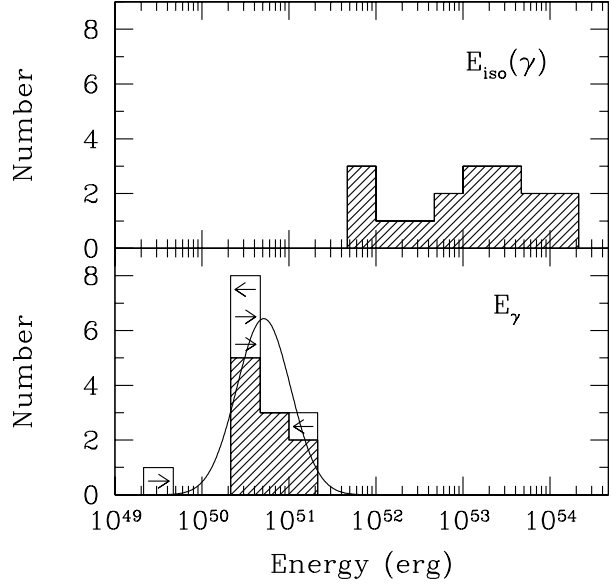}
}
\caption[]{\it The distribution of the apparent isotropic $\gamma$-ray
energy of GRBs with known redshifts (top) versus the
geometry-corrected energy (bottom). While the isotropic energy
$E_{iso}$ spans three orders of magnitudes, the geometrically
corrected energy, $E_\gamma=E_{iso}\theta^2/2$, is very narrowly
distributed.  This implies that the sources of GRBs produce roughly
the same amount of energy, about $5\times 10^{50}$erg, but that energy
is distributed over a variety of angles resulting in a wide
distribution of isotropic energies. From Frail et al. (2001) and Perna
et. al. (2003).
\label{fig:jetsenergy}}
\end{figure}

Frail et al. (2001) collected the jet-break times for a sample
of GRBs with known redshifts. From these, a wide range of jet opening
angles is inferred in GRBs: from 3$^\circ$ to more than 25$^\circ$,
with a strong concentration near 4$^\circ$.  This relatively narrow
collimation implies that the observed GRB rate has to be corrected for
the fact that conical fireballs are visible to only a fraction of
observers. Frail et al. find that the `true' GRB rate is $\sim 500$
times larger than the observed GRB rate.  Although the isotropic
equivalent energies of GRBs range from about 5 $\times$ 10$^{51}$ to
1.4 $\times$ 10$^{54}$ erg, when one corrects the observed
$\gamma$-ray energies for the geometry of the outflow, GRB energies
appear narrowly clustered around $5 \times 10^{50}$ ergs (see
Fig. \ref{fig:jetsenergy}).  Similar conclusions were obtained by Piran et al. (2001) and
Panaitescu \& Kumar (2001b).

The central engines of GRBs thus produce approximately a similar
amount of energy, and the broad range of fluence and luminosity
observed for GRBs appears to be largely the result of a wide variation of
opening angles. The reason why this range in angles exists is
currently not understood.  Our understanding of gamma ray bursts has come a long way in
the past four years.  It is interesting to note that before the
redshift era, most models assumed that the events were standard
candles with energies of about $10^{51}$ erg. As more and more
redshifts were determined, the energy record increased steadily up to
$10^{54}$ erg. The standard candle hypothesis was abandoned.  It is
remarkable that now, when more detailed understanding allows us to
infer the beaming angles of these explosions, the true energy budget
is back at $\sim 10^{51}$ erg, and the explosions are once again
standard candles (though not in the same sense as before).

Postnov, Prokhorov, \& Lipunov (2001), Rossi, Lazzati, \& Rees.(2002), and
Zhang \& M\'esz\'aros (2002) pointed out that another interpretation is
possible for the Frail et. al result. Instead of a variaty of jets
with different opening angles, a standard jet can be invoked with
energy density per unit solid angle falling away from the axis as
$\theta^{-2}$; the differences in the aparent opening angle then come from
variations in the orientation of the observer relative to the jet's
axis. Perna, Sari, \& Frail (2003) showed that the distribution of the observed
opening angles is consistent with this assumption, adding credence to
the universal jet model. If this model is correct, the rate of GRBs is
much lower, because it should not be corrected by the factor of 500 of Frail
et. al.; however, the energy is still low, of order $10^{51}$ ergs.

\section{Polarization - A promising tool}
 
An exciting possibility to further constrain the models and obtain a
more direct proof of the geometrical picture of `jets' is to measure
linear polarization. Varying polarization at optical wavelengths has
been observed in GRB afterglows at the level of a few to ten percent
(Covino et al. 1999, 2002; Wijers et al. 1999; Rol et al. 2000; Bersier et al. 2003).

High levels of linear polarization are usually the smoking gun of
synchrotron radiation.  The direction of the polarization is
perpendicular to the magnetic field and can be as high as $70\%$.
Gruzinov and Waxman (1999) and Medvedev and Loeb (1999)
considered the emission from spherical ejecta which by symmetry should
produce no polarization on the average, except for fluctuations of
order a few percent. Polarization is more natural if the ejecta are a
`jet' and the line of sight to the observer is within the jet but does
not coincide with its axis. In this case, the spherical symmetry is
broken (Gruzinov 1999; Ghisellini \&  Lazzati, 1999; Sari 1999), and the polarization produced by
synchrotron radiation will not vanish. For simplicity, assume
that the magnetic field behind the shock is directed along the shock's
plane (the results hold more generally, as long as the magnetic field
has a preferred direction). The synchrotron polarization from each
part of the shock front, which is perpendicular to the magnetic field,
is therefore directed radially.
 
\begin{figure}
\centerline{
\epsfxsize15pc \epsfbox{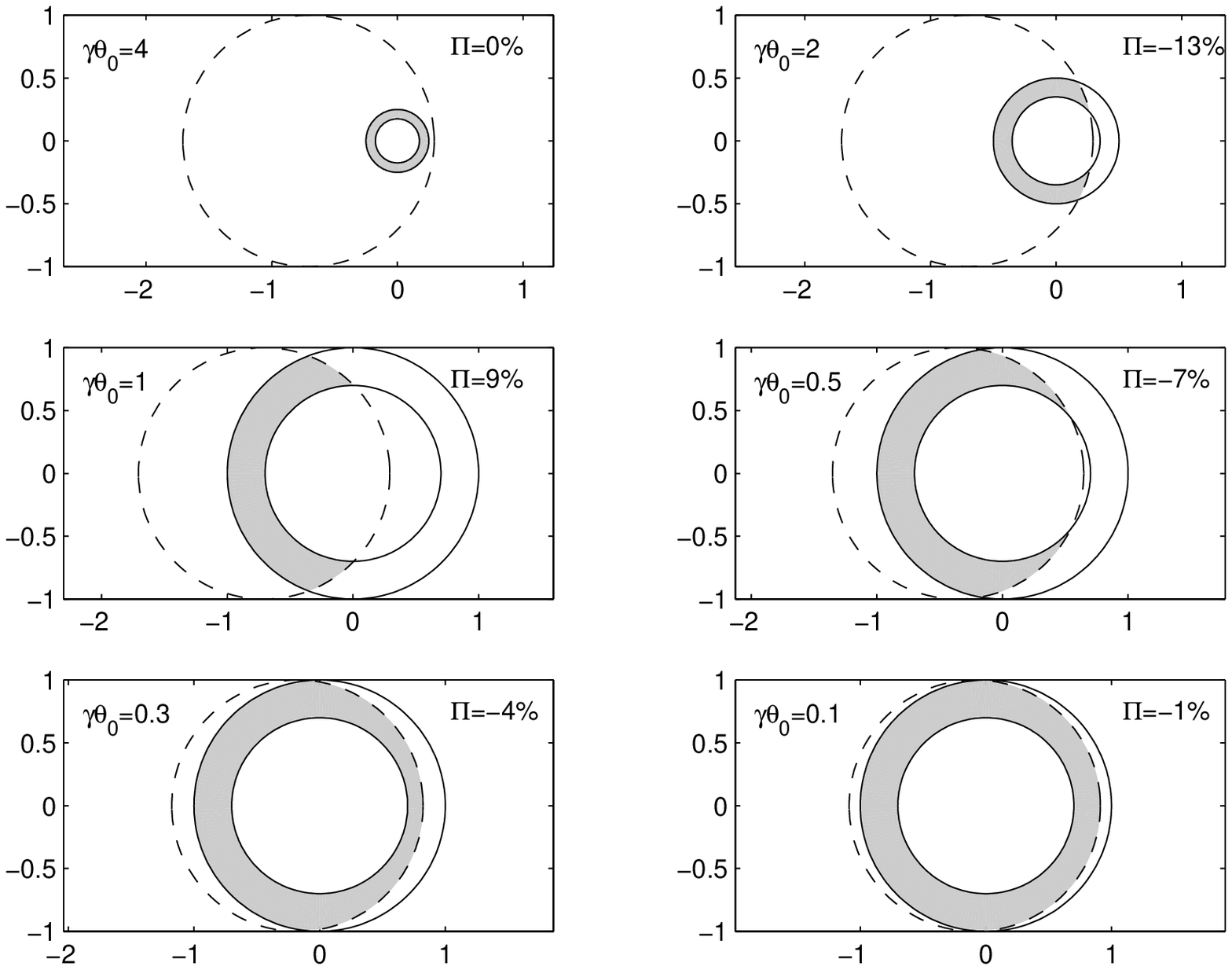} 
\epsfxsize15pc \epsfbox{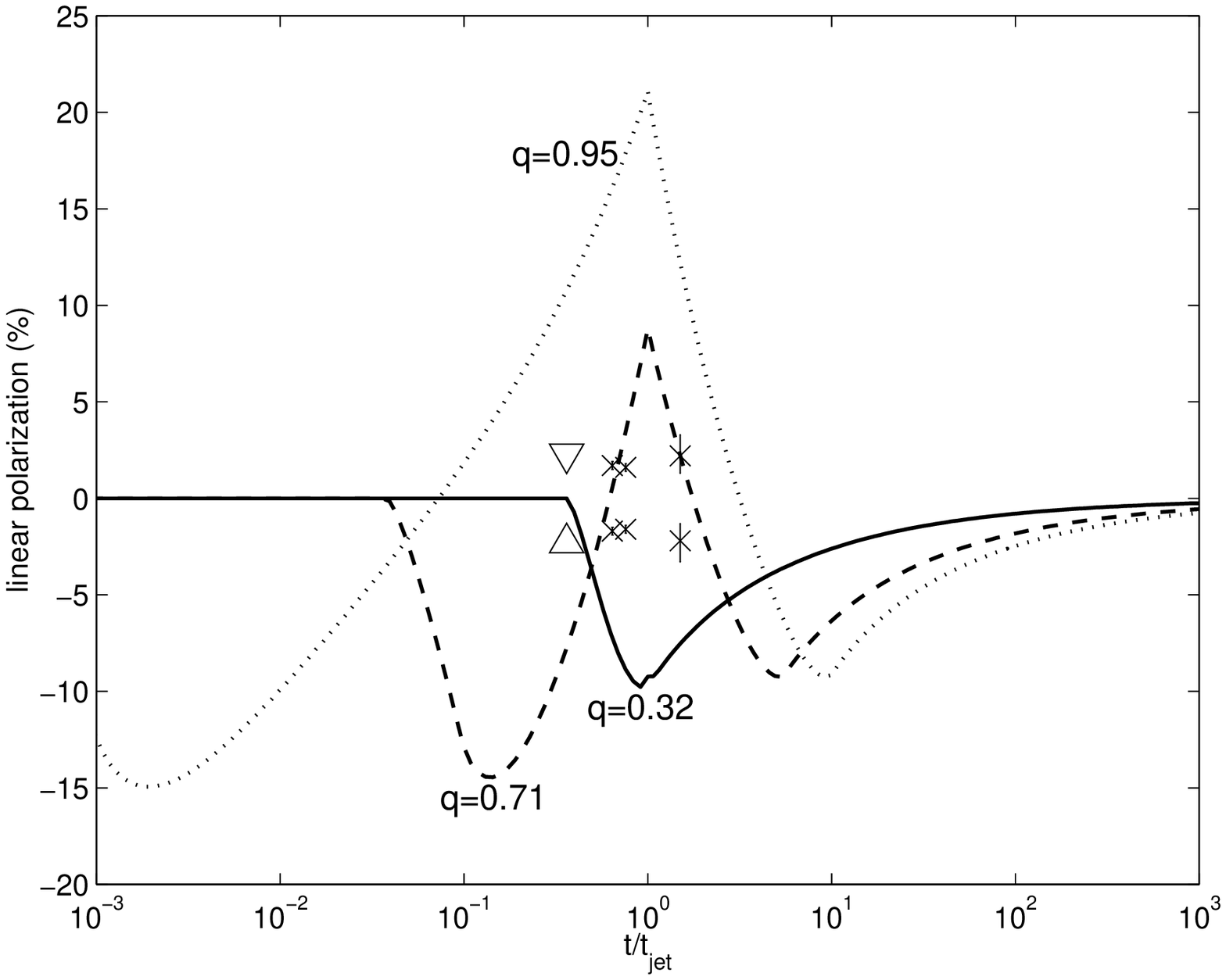}}
\vspace{10pt}
\caption{ Left: Shape of the emitting region.  The dashed line
marks the physical extent of the jet, and solid lines give the
viewable region $1/\gamma$.  The observed radiation arises from the
gray shaded region.  In each frame, the percentage of polarization is
given at the top right and the initial size of the jet relative to
$1/\gamma$ is given on the left.  The frames are scaled so that the
size of the jet is unity.  Right: Observed and theoretical
polarization lightcurves for three possible offsets of the observer
relative to the jet axis Observational data for GRB\,990510 is marked
by crosses (x), assuming $t_{jet}=1.2$\,days.  The upper limit for
GRB\,990123 is given by a triangle, assuming $t_{jet}=2.1$\,days.}
\label{polfig}
\end{figure}

As long as the relativistic beaming angle $1/\gamma$ is narrower than
the physical size of the jet $\theta_0$, one is able to see a full
ring and therefore the radial polarization averages out (the first
frame, with $\gamma\theta_0=4$ of the left plot in Figure
\ref{polfig}). As the flow decelerates, the relativistic beaming angle
$1/\gamma$ becomes comparable to $\theta_0$ and only a part of the
ring is visible; net polarization is then observed. Note that due to
the radial direction of the polarization from each fluid element, the
total polarization is maximal when a quarter ($\gamma\theta_0=2$ in
Figure \ref{polfig}) or when three quarters ($\gamma\theta_0=1$ in
Figure \ref{polfig}) of the ring are missing (or radiate less
efficiently) and vanishes for a full and a half ring. The
polarization, when more than half of the ring is missing, is
perpendicular to the polarization direction when less than half of it
is missing.
 
At late stages the jet expands sideways and since the offset of the
observer from the physical center of the jet is constant, spherical
symmetry is regained. The vanishing and re-occurrence of significant
parts of the ring results in a unique prediction: there should be
three peaks of polarization, with the polarization position angle
during the central peak rotated by $90^{\circ }$ with respect to the
other two peaks. In case the observer is very close to the center,
more than half of the ring is always observed, and therefore only a
single direction of polarization is expected. A few possible
polarization light curves are presented in Figure \ref{polfig}.

\section{The Reverse Shock Emission: Theory and Observations \label{sec:revtheo} \label{sec:revobs} }

The previous sections discussed the theory and the observations of the
`late' afterglow, hours or more after the burst. During that time,
most of the energy of the system was already given to the shocked
surroundings, and it is that region that dominates the
emission. However, during the first few tens of seconds, the evolution
of the Lorentz factor as a function of time is not self-similar. There
are two shocks: a forward shock going into the surrounding medium and
a reverse shock going into the expanding ejecta.
The hydrodynamic details were discussed in Sari \& Piran (1995).


During the initial stages, the internal energy stored behind the
shocked-surrounding matter and the energy of the shocked ejecta are
comparable. However, the temperature of the shocked ejecta is much
lower, typically by a factor of $\gamma \sim 10^2$. This results in an
additional emission component with a typical frequency lower by a
factor of $\gamma^2 \sim 10^4$, which, for typical parameters, is near
the optical passband. Contrary to the `standard' late afterglow, this
emission is very sensitive to the initial Lorentz factor. Theoretical
predictions for such a flash were given in detail by Sari \&
Piran (1999a, 1999c) and were earlier suggested as a possibility by
M\'esz\'aros \& Rees (1997).

\begin{figure}
\centerline{\epsfxsize20pc \epsfbox{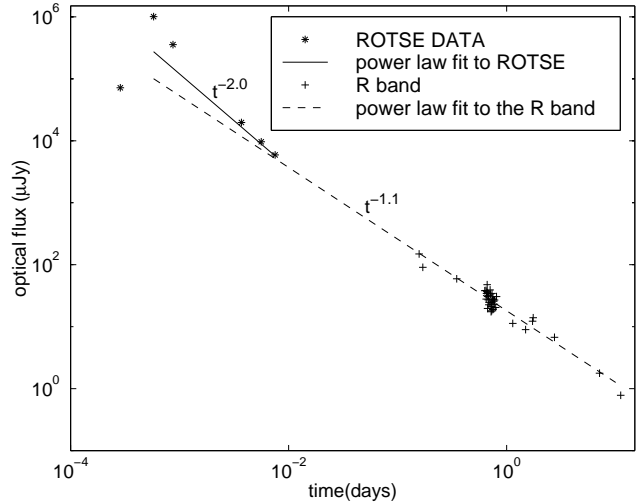}}
\vspace{10pt}
\caption{ R-band light curve of the afterglow of GRB\,990123. The
ROTSE data show that the optical light curve peaked at m$_V \sim 9$
(Akerlof et al. 1999). The dashed line indicates a power law fit to the light
curve (for $t > 0.1$ days), which has exponent $-1.12 \pm 0.03$ (from
Galama et al. 1999).
\label{fig:0123}}
\end{figure}

One of the most exciting events in the field of afterglow studies was
the detection of bright (9th magnitude) optical emission
simultaneous with GRB\,990123 by the ROTSE team (Akerlof et al. 1999). The ROTSE
telescope obtained its first images only 22 seconds after the start of
GRB\,990123 (i.e. during the GRB), following a notification received
from BATSE aboard the Compton Gamma-Ray Observatory. The ROTSE observations
show that the optical light curve peaked at m$_V \sim 9$ magnitudes
some 60 seconds after the event (Akerlof et al. 1999). After maximum a fast
decay followed for at least 15 minutes. The late-time afterglow
observations show a more gradual decline
(Galama et al. 1999; Kulkarni et al. 1999a; Castro-Tirado et al. 1999; Fruchter et al. 1999, Sari \& Piran 1999b) (see Fig. \ref{fig:0123}).

The redshift $z = 1.6$, inferred from absorption features in the OT's
spectrum, implies that the optical flash would have been as bright as
the full moon had the GRB occurred in the nearby galaxy M31
(Andromeda). A different way to put this in perspective is that the
flash was some 18 mag brighter (absolute) than the brightest
supernovae. Galama et al.  (1999) have shown that if one
assumes that the emission detected by ROTSE comes from a
non-relativistic source of size $ct$, that then the observed
brightness temperature $T_b\ge 10^{17}$\,K of the optical flash
exceeds the Compton limit of $10^{12}$\,K. This confirms the highly
relativistic nature of the GRB source.

The observed optical properties of this event are well described by
emission from the reverse shock that initially decelerates the ejecta,
provided that the initial Lorentz factor is about 200 (Sari \& Piran 1999b; M\'esz\'aros \& Rees 1999). It
takes tens of seconds for the reverse shock to sweep through the
ejecta and produce the bright flash. Later, the shocked hot matter
expands adiabatically and the emission quickly shifts to lower
frequencies and considerably weakens.

The ROTSE observations show that the prompt optical and $\gamma$-ray
light curves do not track each other (Akerlof et al. 1999). In addition,
detailed comparison of the prompt optical emission with the BATSE
spectra of GRB\,990123 (at three epochs for which both optical and
gamma-ray information is available) shows that the ROTSE emission is
not a simple extrapolation of the GRB spectrum to much lower energies
(Galama et al. 1999; Briggs et al. 1999). 

If this interpretation is correct, GRB\,990123 would be the first
burst in which all three emitting regions have been seen: internal
shocks causing the GRB, the reverse shock causing the prompt optical
flash, and the forward shock causing the afterglow.  The emissions
thus arise from three different emitting regions, explaining the lack
of correlation between the GRB, the prompt optical and the late-time
optical emission (Galama et al. 1999) (but see Liang et al. 1999).

Another new ingredient that was found in GRB\,990123 is a radio flare
(Kulkarni et al. 1999b). Contrary to all other afterglows, where the radio peaks
around a few weeks and then decays slowly, this burst had a fast
rising flare, peaking around a day and then decaying quickly. This can
be interpreted as emission from the cooling ejecta that was earlier on
heated by the reverse shock. Using the Blandford and McKee (1976)
self-similar solution to derive the evolution of the ejecta and its
emission properties one finds that the typical frequency scales as
$\nu_{ m}^r \propto t^{-73/48}$ and the flux at that frequency scales
as $F_{ m}^r \propto t^{-47/48}$ (Sari \& Piran 1999a) (see Kobayashi \& Sari 2001) for
revised scalings when the temperature of the ejecta is non-
relativistic). Therefore, within a day the emission from the
adiabatically cooling ejecta that produced the $60$s optical flash in
GRB\,990123 is expected to shift to radio frequencies
(Sari \& Piran 1999b). Using the observed optical flash and the above
scalings, a good fit to the radio data is obtained.  The optical flash
and the radio flare may therefore be related.

Given the above interpretation of the reverse shock emission, it is
important to ask whether GRB\,990123 is an exception, or whether the
phenomena of radio flares and optical flashes is more common. Radio
flares appear to exist in other cases (Frail et al. 2001).  However, since
early radio data is usually sparse, and these events did not have an
early optical observation to find the associated optical flash, the
interpretation in terms of emission from the reverse shock is less
secure than in the case of GRB\,990123.  In the optical, from robotic
optical experiments such as ROTSE and LOTIS, strong upper limits
exist for several bursts.  The upper limits show that the optical
flash does not scale with the fluence of the event
(Akerlof et al. 2000; Kehoe et al. 2001). However, with reasonably small changes in the
density or the initial Lorentz factor, those events could have escaped
detection (Kobayashi 2000). HETE-II, or future satellites like Swift,
provide or will provide accurate positioning on timescales of seconds, and strong
constraints on the generality of optical flashes and radio flares will
be obtained.



\section{GRB Host Galaxies and Redshifts}

Host galaxies of GRBs serve a dual purpose:  they determine the redshifts,
which are necessary for a complete physical modeling of the bursts,
and they provide some insights about the possible nature of the progenitors,
e.g., their relation to massive star formation, etc.  The subject has been
reviewed previously, e.g., by Djorgovski et al. (2001b, 2002).

\subsection{Overall Properties of GRB Hosts}

\begin{figure}
\centering
\centerline{\epsfxsize15pc\epsfbox{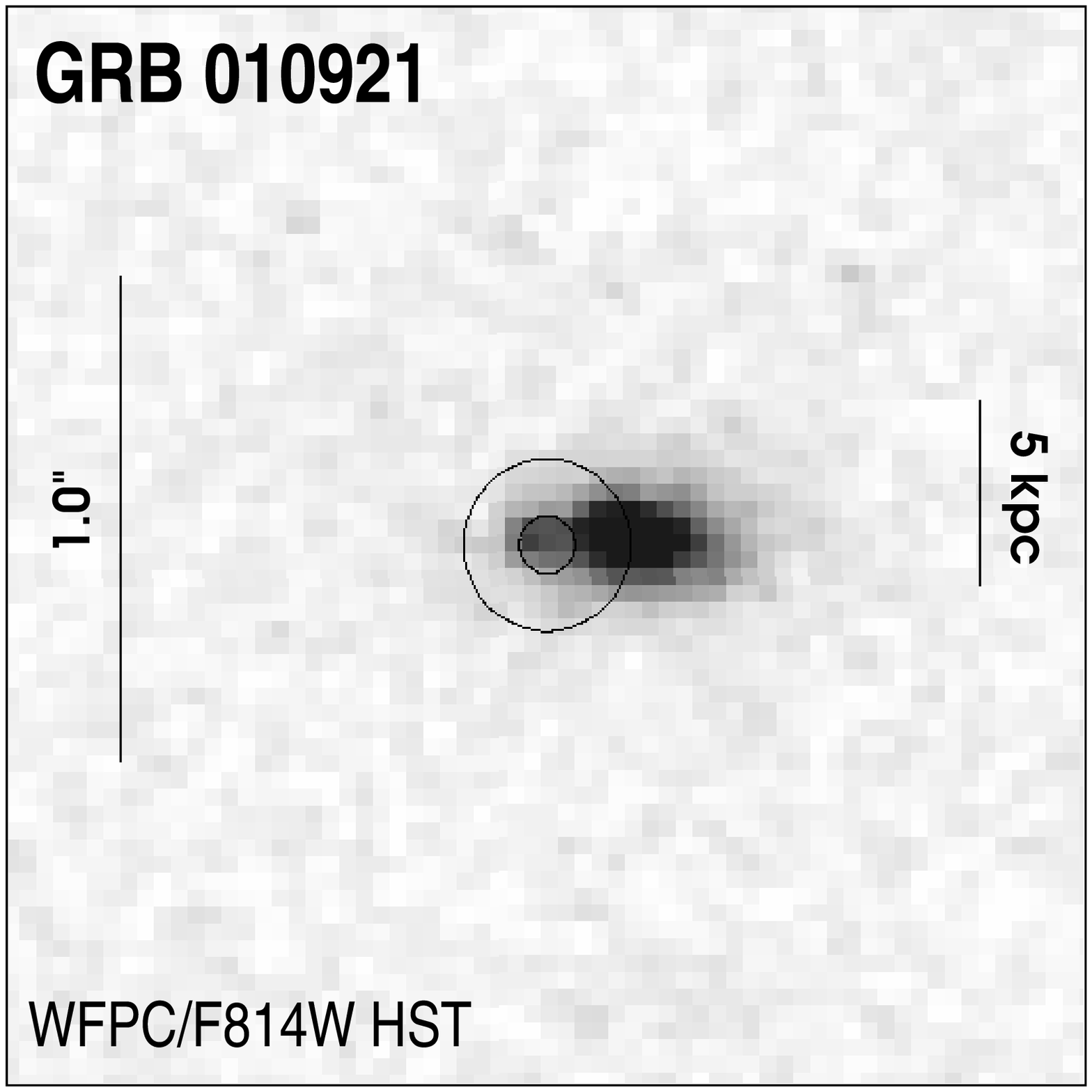}\ \ \epsfxsize14.5pc \epsfbox{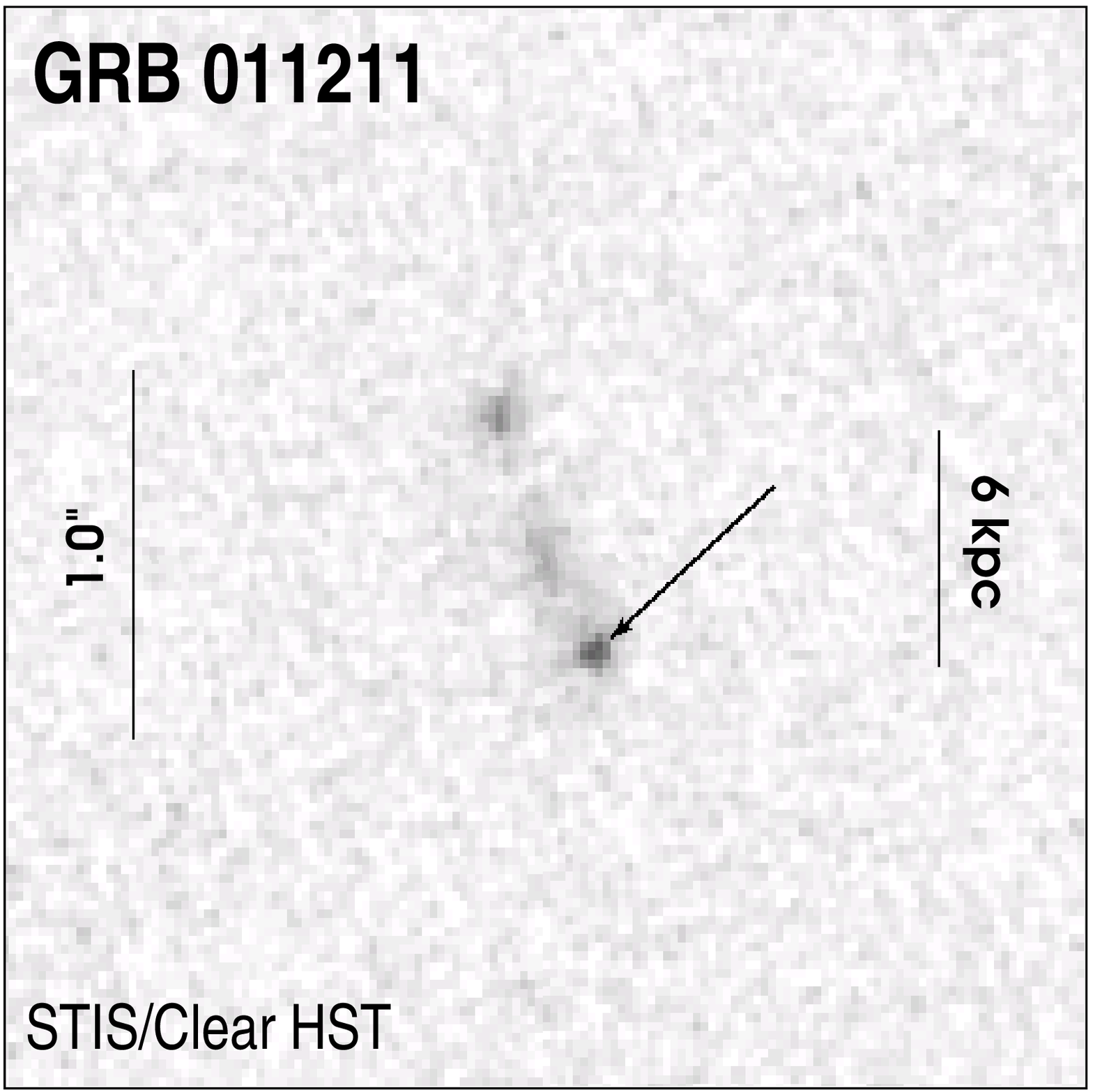}}
\caption{Typical HST images of two GRB host galaxies.  The OTs are indicated by
the circle (for GRB 010921) and by the arrow (for GRB 011211).  Side bars
give the projected angular scale, and the physical scale at the source.
From Bloom, Kulkarni, \& Djorgovski (2002), images courtesy of J.S. Bloom.}
\label{fig:galaxies}
\end{figure}

As of this writing ($\sim$ late 2002), plausible or certain host galaxies have
been found for all but 1 or 2 of the bursts with optical, radio, or x-ray
afterglows localised with arcsecond precision.  Two examples are shown in
figure
\ref{fig:galaxies}.  The median apparent
magnitude is $R \approx 25$ mag, with tentative detections or upper limits
reaching down to $R \approx 29$ mag.  The missing cases are at least
qualitatively consistent with being in the faint tail of the observed
distribution of host galaxy magnitudes.

Down to $R \sim 25$ mag, the observed distribution is consistent with deep
field galaxy counts (Brunner, Connolly, \& Szalay 1999),
but fainter than that, complex selection effects may be playing a role.
It can also be argued that the observed distribution should correspond
roughly to luminosity-weighted field galaxy counts.  However, the actual
distribution would depend on many observational selection and physical
(galaxy evolution) effects, and a full interpretation of the observed
distribution of GRB host galaxy magnitudes requires a careful modeling.
We note also that the observations in the visible probe the
UV in the restframe, and are thus especially susceptible to extinction. 
However, sub-mm detections of dusty GRB hosts are currently limited by the
available technology to only a handful of ultraluminous sources.

\begin{figure}[h]
\centering
\epsfxsize25pc
\epsfbox{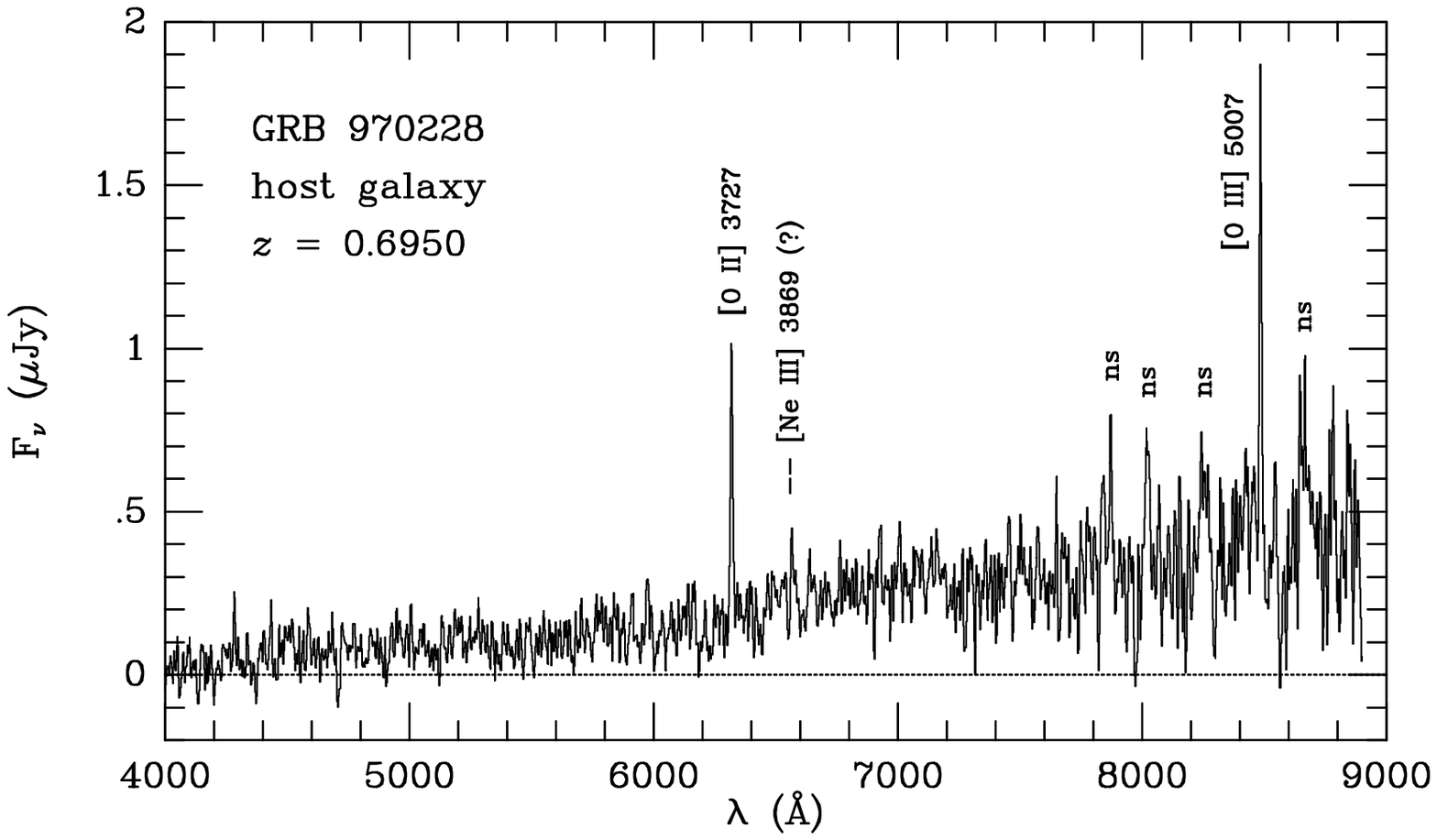}
\epsfxsize25pc 
\epsfbox{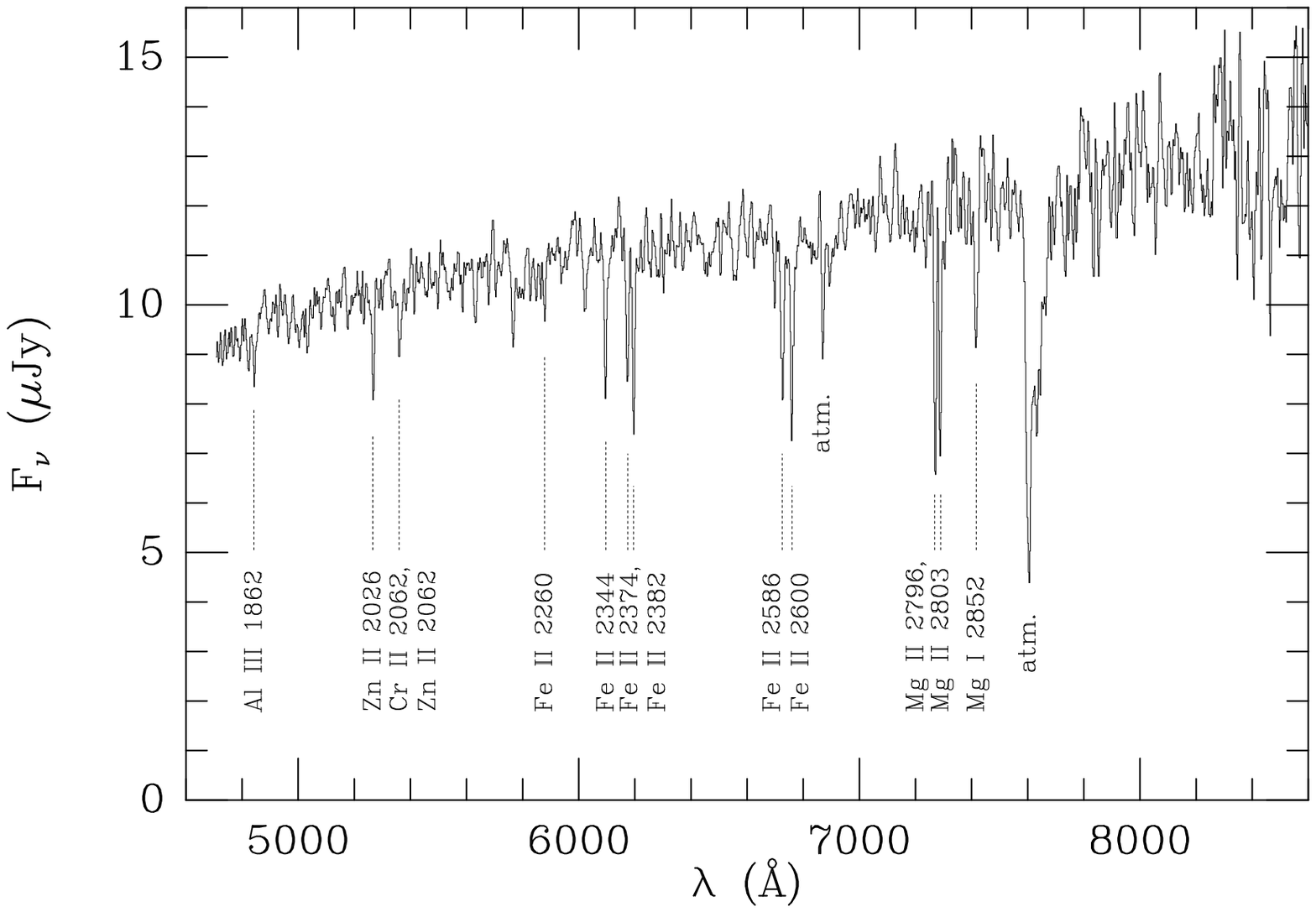}
\caption{Typical spectra of a GRB host galaxy, showing the standard emission lines
indicative of active star formation (for GRB 970228, top), and of an OT, showing
the strong absorption lines from the ISM in the host galaxy (for GRB 990123, bottom).
The top figure is from Bloom, Djorgovski \& Kulkarni (2001).  The bottom figure
is  reprinted by permission from Nature (Kulkarni et al.1999a) copyright 1999
Macmillan Publishers Ltd.}
\label{fig:spectra}
\end{figure}

Starting with the first redshift measurement which unambiguosly demonstrated
the cosmological nature of GRBs (Metzger et al. 1997)
there are now (late 2002) over 30 redshifts measured for GRB hosts 
and/or afterglows.  The median redshift is $\langle z \rangle \approx 1.0$,
spanning the range from 0.25 (or 0.0085, if the association of GRB 980425
with SN 1998bw is correct)
to 4.5 (for GRB 000131).
The majority of redshifts so far are from the spectroscopy of host galaxies,
but an increasing number are based on the absorption-line systems seen in
the spectra of the
afterglows (which are otherwise featureless power-law continua).  Figure \ref{fig:spectra} shows
two examples.  Reassuring
overlap exists in several cases; invariably, the highest-$z$ absorption system
corresponds to that of the host galaxy, and has the strongest lines.
In some cases (a subset of the so-called ``dark bursts'') no optical
transient (OT) is detected, but a combination of the X-ray (XT) and radio
transient (RT) unambiguously pinpoints the host galaxy.

A new method for obtaining redshifts may come from the X-ray spectroscopy of
afterglows, using the Fe K line at $\sim 6.55$ keV (Piro et al. 1999, 2000; Antonelli et al. 2000),
or the Fe absorption edge at $\sim 9.28$ keV (Weth et al. 2000; Yohshida et al. 1999; Amati et al. 2000).
Rapid X-ray spectroscopy of GRB afterglows may become a powerful
tool for understanding their physics and origins. 

Are the GRB host galaxies special in some way?  If GRBs are somehow related to
massive star formation (e.g., Paczy\'nski 1998, Totani, 1997, etc.), it may be
worthwhile to examine their absolute luminosities and star formation rates
(SFR), or spectroscopic properties in general.  This is hard to answer
(Krumholz, Thorsett, \& Harrison 1998; Hogg \& Fruchter 1999; Schaefer 2000) 
from their visible ($\sim$ restframe UV) luminosities alone: the observed 
light traces an indeterminate mix of recently formed stars and an older
population, and cannot be unambiguously interpreted in terms of either the total
baryonic mass, or the instantaneous SFR.  

The magnitude and redshift distributions of GRB host galaxies are typical for
the normal, faint field galaxies, as are their morphologies
(Odewahn et al. 1998; Holland 2001; Bloom, Kulkarni \& Djorgovski 2002)
when observed with the HST: they are often compact, and sometimes suggestive of a merging
system (Djorgovski, Bloom \& Kulkarni 2001; Hjorth et al. 2002),
but that is not unusual for galaxies at comparable redshifts. 

Within the host galaxies, the distribution of GRB-host offsets follows 
the light distribution closely (Bloom, Kulkarni \& Djorgovski 2002), 
which is roughly proportional to the density of star formation (especially for
the high-$z$ galaxies).  It is thus fully consistent with a progenitor
population associated with the sites of massive star formation.

Spectroscopic measurements provide direct estimates of recent, massive
SFR in GRB hosts.  Most of them are based on 
the luminosity of the [O II] 3727 doublet (Kennicut 1998), 
the luminosity of the UV continuum at $\lambda_{rest} = 2800$ \AA\ (Madau, Pozzetti, \& Dickinson 1998),
in some cases (e.g., Kulkarni et al. 1998)
from the luminosity of the Ly$\alpha$ 1216 line,
and in others (e.g., Djorgovski et al. 1998)
from the luminosity of Balmer lines (Kennicut 1998).
All of these estimators are susceptible to the internal extinction and its
geometry, and have an intrinsic scatter of at least 30\%.
The observed $unobscured$ SFR's range from a few tenths to a few 
$M_\odot$ yr$^{-1}$.  Applying the reddening corrections derived from the
Balmer decrements of the hosts, or from the modeling of the broad-band colors
of the OTs (and further assuming that they are representative of the mean
extinction for the corresponding host galaxies) increases these numbers
typically by a factor of a few.  All this is entirely 
typical for the normal field galaxy population at comparable redshifts.
However, such measurements are completely insensitive to any fully obscured
SFR components.

Equivalent widths of the [O II] 3727 doublet in GRB hosts, which may provide
a crude measure of the SFR per unit luminosity (and a worse measure of the
SFR per unit mass), are on average somewhat higher (Djorgovski et al. 2001a)
than those observed in magnitude-limited field galaxy samples at comparable
redshifts (Hogg et al. 1998).
A larger sample of GRB hosts, and a good comparison sample, matched both 
in redshift and magnitude range, are necessary before any solid conclusions can
be drawn from this apparent difference. 

One intriguing hint comes from the flux ratios of [Ne III] 3869 to
[O II] 3727 lines: they are on average a factor of 4 to 5 higher in GRB
hosts than in star forming galaxies at low redshifts (Djorgovski et al. 2001b). 
Strong [Ne III] 
requires photoionization by massive stars in hot H II regions, and may represent
indirect evidence linking GRBs with massive star formation.

The interpretation of the luminosities and observed star formation rates is
vastly complicated by the unknown amount and geometry of extinction.  The
observed quantities (in the visible) trace only the unobscured stellar
component, or the components seen through optically thin dust.  Any 
stellar and star formation components hidden by optically thick dust cannot
be estimated at all from these data, and require radio and sub-mm observations.

Both observational windows, the optical/NIR (rest-frame UV) and the sub-mm
(rest-frame FIR)suffer from some biases: the optical band is significantly
affected by dust obscuration, while the sub-mm and radio bands lack
sensitivity, and therefore uncover only the most prodigiously star-forming
galaxies.  As of late 2002, radio and/or sub-mm emission powered by
obscured star formation has been detected from 4 GRB hosts
(Berger, Kulkarni \& Frail 2001; Berger et al. 2002b; Frail et al. 2002).  The surveys to date are sensitive only
to the ultra-luminous ($L > 10^{12} L_\odot$) hosts, with SFR of several
hundred M$_\odot$ yr$^{-1}$.  Modulo the uncertainties posed by the small
number statistsics, the surveys indicate that about 20\% of GRB hosts are
objects of this type, where about 90\% of the total star formation takes
place in obscured regions.

Given the uncertainties of the geometry of optically thin and optically thick
dust, optical colors of GRB hosts cannot be used to make any
meaningful statements about their net star formation activity.  The broad-band
optical colors of GRB hosts are not distinguishable from those of normal
field galaxies at comparable magnitudes and redshifts (Bloom, Djorgovski, \& Kulkarni 2001; Sokolov et al. 2001).
It is notable that the optical/NIR colors of GRB hosts detected in the 
sub-mm are much bluer than typical sub-mm selected galaxies, suggesting that
the GRB selection may be probing a previously unrecognised population of
dusty star-forming galaxies.

On the whole, the GRB hosts seem to be representative of the normal,
star-forming
field galaxy population at comparable redshifts, and so far there is no
evidence for any significant systematic differences between them.

\subsection{GRB Hosts in the Context of Galaxy Evolution}

The observed redshift distribution of GRB hosts is about what is expected for
an evolving, normal
field galaxy population at these magnitude levels.  There is an excellent
qualitative correspondence between the observations and simple galaxy evolution
models (Mao \& Mo 1998).  

If GRB's follow the luminous mass, then the expected
distribution would be approximated by the luminosity-weighted galaxy luminosity
function (GLF) for the appropriate redshifts.
The hosts span a wide range of luminosities, with a characteristic absolute
restframe B band magnitude $M_{B,*} \approx -20$ mag, approximately half a
magnitude fainter than in the GLF at $z \approx 0$,
but comensurate with the late-type (i.e., star forming disk) galaxy
population at $z \approx 0$ (Madgwick et al. 2002; Norberg et al. 2002).  
This is somewhat surprising, since one
expects that the evolutionary effects would make the GRB host  galaxies,
with a typical $z \sim 1$, brighter than their descendants today.  The GRB
host GLF also has a somewhat steeper tail than the composite GLF at 
$z \approx 0$, but again similar to that of the star-forming, late-type
galaxies.  This is in a broad agreement with the results of deep redshift
surveys which probe the evolution of field galaxy populations out to $z \sim 1$
(Lilly et al. 1995; Ellis 1997; Fried et al. 2001; Lin et al. 1999).

The interpretation of these results is complex: the observed light reflects an
unknown combination of the unobscured fraction of recent star formation
(especially in the high-$z$ galaxies, where we observe the restframe UV
continuum) and the stellar populations created up to that point.  Our 
understanding of the field galaxy evolution in the same redshift range as
probed by the GRB hosts is still largely incomplete.  Different selection
effects may be plaguing the field and the GRB host samples.  While much
remains to be done, it seems that GRB hosts provide a new, independent
check on the traditional studies of galaxy evolution at moderate and high
redshifts.

\section{GRBs and Cosmology}

While interesting on their own, GRBs are now rapidly becoming powerful tools to
study the high-redshift universe and galaxy evolution, thanks to their apparent
association with massive star formation, and their brilliant luminosities.

There are three basic ways of learning about the evolution of luminous matter
and gas in the universe.  First, a direct detection of sources (i.e., galaxies)
in emission, either in the UV/optical/NIR (the unobscured components), or
in the FIR/sub-mm/radio (the obscured component).  Second, the detection of
galaxies selected in absorption along the lines of sight to luminous background
sources, traditionally QSOs.  Third, diffuse extragalactic backgrounds, which
bypass all of the flux or surface brightness selection effects plaguing all
surveys of discrete sources found in emission, but at a price of losing the
redshift information, and the ability to discriminate between the luminosity
components powered by star formation and powered by AGN.  Studies of GRB hosts
and afterglows can contribute to all three of these methodological approaches,
bringing in new, independent constraints for models of galaxy evolution and
of the history of star formation in the universe.

\subsection{Dark Bursts: Probing the Obscured Star Formation History}

Already within months of the first detections of GRB afterglows, no OT's were
found associated with some well-localised bursts despite deep and rapid
searches; the prototype ``dark burst'' was GRB 970828 (Djorgovski et al. 2001a).
Perhaps the most likely explanation for the non-detections of OT's when
sufficiently deep and prompt searches are made is that they are obscured by
dust in their host galaxies.  This is an obvious culprit if indeed GRBs are
associated with massive star formation.

Support for this idea also comes from detections of RTs without OTs, including
GRB 970828, 990506, and possibly also 981226 (see Frail et al. 2000 and Taylor et al. 2000).
Dust reddening has been detected directly in some OTs 
(e.g., Ramakaprash et al. 1998; Bloom et al. 1998; Djorgovski et al. 1998, etc.);
however, this only covers OTs seen through optically thin dust, and there
must be others, hidden by optically thick dust.
An especially dramatic case was the RT (Taylor et al. 1998) and 
IR transient (Larkin et al. 1998)
associated with GRB 980329 (Yost et al. 2002).
We thus know that at least some GRB OTs must be obscured by dust. 

The census of OT detections for well-localised bursts can thus provide a
completely new and independent estimate of the mean obscured star formation
fraction in the universe.  Recall that GRBs are now detected out to 
$z \sim 4.5$ and that there is no correlation of the observed fluence with 
the redshift (Djorgovski et al. 2002),
so that they are, at least to a first approximation, good probes of the star
formation over the observable universe.  

As of late 2002, there have been $\sim 70$ adequately deep and rapid
searches for OTs from well-localised GRBs. 
We define ``adequate searches'' as reaching at least to $R \sim 20$ mag within
less than a day from the burst, and/or to at least to $R \sim 23 - 24$ mag
within 2 or 3 days; this is a purely heuristic, operational definition,
and an intentionally liberal one.  
In just over a half of such searches, OTs were found.
Inevitably, some OTs may have been missed due to an intrinsically low flux,
an unusually rapid decline rate (Fynbo et al. 2001; Berger et al. 2002a), 
or very high redshifts (so that the brightness in the commonly used $BVR$ bands
would be affected by the intergalactic absorption).
Thus the $maximum$ fraction of all OTs (and therefore massive star formation)
hidden by the dust is $\sim 50$\%.

This is a remarkable result.  It broadly agrees with the estimates that there
is roughly an equal amount of energy in the diffuse optical and FIR backgrounds
(see, e.g., Madau 1999).  This is contrary to some claims in the literature
which suggest that the fraction of the obscured star formation was much higher
at high redshifts.  Recall also that the fractions of the obscured and
unobscured star formation in the local universe are comparable.  

There is one possible loophole in this argument: GRBs may be able to destroy
the dust in their immediate vicinity (up to $\sim 10$ pc?) (Waxman \& Draine 2000; Galama \& Wijers 2000),
and if the rest of the optical path through their hosts ($\sim$ kpc scale?)
was dust-free, OTs would become visible.  Such a geometrical arrangement may
be unlikely in most cases, and our argument probably still applies.
A more careful treatment of the dust evaporation geometry is needed, but
it is probably safe to say that GRBs can provide a valuable new constraint
on the history of star formation in the universe.

\subsection{GRBs as Probes of the ISM in Evolving Galaxies}

Absorption spectroscopy of GRB afterglows is now becoming a powerful new
probe of the ISM in evolving galaxies, complementary to the traditional studies
of QSO absorption line systems.  The key point is that the GRBs almost by
definition (that is, if they are closely related to the sites of
ongoing or recent massive star formation, as the data seem to indicate)
probe the lines of sight to dense, central regions of their host galaxies
($\sim 1 - 10$ kpc scale).  On the other hand, the QSO absorption systems
are selected by the gas cross section, and favor large impact parameters
($\sim 10 - 100$ kpc scale), mostly probing the gaseous halos of field galaxies,
where the physical conditions are very different.

The growing body of data on GRB absorption systems shows exceptionally high
column densities of gas, when compared to the typical QSO absorption systems;
only the highest column density DLA systems (themselves ostensibly star-forming
disks or dwarfs) come close (Savaglio, Fall, \& Fiore 2002; Castro et al. 2002; Mirabal et al. 2002).
This is completely consistent with the general picture described above.
(We are refering here to the highest redshift absorbers seen in the afterglow
spectra, which are presumably associated with the host galaxies themselves;
lower redshift, intervening absorbers are also frequently seen, and their
properties appear to be no different from those of the QSO absorbers.)

This opens the interesting prospect of using GRB absorbers as a new probe of
the chemical enrichment history in galaxies in a more direct fashion than
what is possible with the QSO absorbers, where there may be a very complex
dynamics of gas ejection, infall, and mixing at play.  

Properties of the GRB absorbers are presumably, but not necessarily (depending
on the unknown geometry of the gas along the line of sight) reflecting the ISM
of the circum-burst region.  Studies of their chemical composition do not yet
reveal any clear anomalies, or the degree of depletion of the dust, but the
samples in hand are still too small to be really conclusive.  Also, there
have been a few searches for the variability of the column density of the gas
on scales of hours to days after the burst, with no clear detections so far.
Such an effect may be expected if the burst afterglow modifies the physical
state of the gas and dust along the line of sight by the evaporation of the
dust grains, additional photoionization of the gas, etc.  However, it is
possible that all such changes are observable only on very short time scales,
seconds to minutes after the burst.  In any case, a clear detection of a
variable ISM absorption against a GRB afterglow would be a very significant
result, providing new insight into the cisrcumstances of GRB origins.

\subsection{High-Redshift GRBs: Probing the Primordial Star \\
Formation and Reionization}

Possibly the most interesting use of GRBs in cosmology is as probes of the
early phases of star and galaxy formation, and the resulting reionization of
the universe at $z \sim 6 - 20$.  If GRBs reflect deaths of massive stars,
their very existence and statistics would provide a superb probe of the
primordial massive star formation and the initial mass function (IMF).  They would be by far the most 
luminous sources in existence at such redshifts (much brighter than SNe, and
most AGN), and they may exist at redshifts where there were $no$ luminous
AGN.  As such, they would provide unique new insights into the physics and
evolution of the primordial IGM during the reionization era (see, e.g.,
Lamb \& Reichart 2001; Loeb 2002a,b).

There are two lines of argument in support of the existence of copious numbers
of GRBs at $z > 5$ or even 10.  First, a number of studies using photometric
redshift indicators for GRBs suggests that a substantial fraction (ranging
from $\sim 10$\% to $\sim 50$\%) of all bursts detectable by past,
current, or forthcoming missions may be originating at such high redshifts,
even after folding in the appropriate spacecraft/instrument selection
functions (Fenimore \& Ramirez-Ruiz 2002; Reichart et al. 2001; Lloyd-Ronning, Fryer, \& Ramirez-Ruiz 2002).

Second, a number of modern theoretical studies suggest that the very first
generation of stars, formed through hydrogen cooling alone, were very
massive, with $M \sim 100 - 1000 ~M_\odot$ (Bromm, Coppi \& Larson 1999;
Abel, Bryan, \& Norman 2000; Bromm, Kudritzki, \& Loeb 2001; Bromm, Coppi \& Larson 2002;
Abel, Bryan \& Norman 2002).
While it is not yet absolutely clear that some as-yet unforseen effect would
lead to a substantial fragmentation of a protostellar object of such a mass, a
top-heavy primordial IMF is at least plausible.  It is also not yet 
completely clear that the (probably spectacular) end of such an object
would generate a GRB, but that too is at least plausible (Fryer, Woosley \& Heger 2001).
Thus, there is some real hope that significant numbers of GRBs and their
afterglows would be detectable in the redshift range $z \sim 5 - 20$,
spanning the era of the first star formation and cosmic reionization (Bromm \& Loeb 2002).

Spectroscopy of GRB aftergows at such redshifts would provide a crucial,
unique information about the physical state and evolution of the primordial
ISM during the reionization era.  The end stages of the cosmic reionization
have been detected by spectroscopy of QSOs at $z \sim 6$
(Djorgovski et al. 2001c; Fan et al. 2001; Becker et al. 2001).
GRBs are more useful in this context than the QSOs, for several reasons.
First, they may exist at high redshifts where there were no comparably
luminous AGN yet.  Second, their spectra are highly predictable power-laws,
without complications caused by the broad Ly$\alpha$ lines of QSOs, and can
reliably be extrapolated blueward of the Ly$\alpha$ line.  Finally, they would
provide a genuine snapshot of the intervening ISM, without an appreciable
proximity effect which would inevitably complicate the interpretation of
any high-$z$ QSO spectrum (luminous QSOs excavate their Stromgren spheres
in the surrounding neutral ISM out to radii of at least a few Mpc, whereas
the primordial GRB hosts would have a negligible effect of that type;
see, e.g., Lazzati et al.(2001).

Detection of high-$z$ GRBs is thus an urgent cosmological task.  It requires
a rapid search for afterglows, as well as high-resolution
follow-up spectroscopy, in both the optical and NIR.  However, such effort would
be well worth the considerable scientific rewards in the end.

\section{Acknowledgments}

SGD and RS wish to thank numerous collaborators, including
S.R. Kulkarni,
D.A. Frail,
F.A. Harrison,
J.S. Bloom,
T. Galama,
D. Reichart,
D. Fox,
E. Berger,
P. Price,
S. Yost,
A. Soderberg,
S.M. Castro,
A. Mahabal,
R. Goodrich,
F. Chaffee,
J. Halpern,
and many others.
Our work was supported by grants from the NSF, NASA, and private donors.

\end{document}